\documentclass[a4paper,11pt]{article}
\pdfoutput=1 

\usepackage{enumerate} 
\usepackage{afterpage} 
\usepackage{amsmath} 
\usepackage{jcappub} 

\usepackage[T1]{fontenc} 
\usepackage[usenames,dvipsnames]{xcolor}

\newcommand{\ex}{t\,$\times$\,y}
\newcommand{\high}[1]{$^{\textnormal{\tiny #1}}$}
\newcommand{\leff}{${\cal L}_\textnormal{\footnotesize eff}$}

\title{\boldmath Dark matter sensitivity of multi-ton liquid xenon detectors}

\author[a,1]{Marc Schumann,\note{Corresponding author.}}
\author[b]{Laura Baudis,}
\author[a]{Lukas B\"utikofer,}
\author[b]{\\ Alexander Kish,}
\author[c]{Marco Selvi}

\affiliation[a]{Albert Einstein Center for Fundamental Physics, Universit\"at Bern, Switzerland}
\affiliation[b]{Physik-Institut, Universit\"at Z\"urich, Switzerland}
\affiliation[c]{INFN Bologna, Italy}

\emailAdd{marc.schumann@lhep.unibe.ch}
\emailAdd{lbaudis@physik.uzh.ch}
\emailAdd{lukas.buetikofer@lhep.unibe.ch}
\emailAdd{alexkish@physik.uzh.ch}
\emailAdd{marco.selvi@bo.infn.it}

\abstract{We study the sensitivity of multi ton-scale time projection chambers using a liquid xenon target, e.g., the proposed DARWIN instrument, to spin-independent and spin-dependent WIMP-nucleon scattering interactions. Taking into account realistic backgrounds from the detector itself as well as from neutrinos, we examine the impact of exposure, energy threshold, background rejection efficiency and energy resolution on the dark matter sensitivity. With an exposure of 200\,\ex \ and assuming detector parameters which have been already demonstrated experimentally, spin-independent cross sections as low as $2.5 \times 10^{-49}$\,cm$^2$ can be probed for WIMP masses around 40\,GeV/$c^2$. Additional improvements in terms of background rejection and exposure will further increase the sensitivity, while the ultimate WIMP science reach will be limited by neutrinos scattering coherently off the xenon nuclei.}

\begin{document}
\maketitle
\flushbottom

\section{Introduction}

The nature of dark matter, contributing about 27\% to the matter and energy content of our Universe~\cite{ref::planck}, is one of the outstanding open questions in physics. A promising dark matter candidate is the weakly interacting massive particle (WIMP)~\cite{ref::wimpmiracle}, which is searched for in various experiments world-wide. These projects aim at its direct detection by measuring the nuclear recoil signals left by WIMPs scattering in sensitive detectors operated underground~\cite{ref::dmreview}. Dual-phase time projection chambers (TPCs) filled with liquid xenon (LXe)~\cite{ref::doublephase,ref::2phase_chepel,ref::2phase_schumann} measure the light (S1) and the charge signal (S2) generated by a particle interaction. A drift field across the TPC removes the ionization charges from the interaction site for detection. The combination of both channels allows for the reconstruction of the event vertex and its multiplicity, and is used to distinguish signal-like nuclear recoil (NR) events, induced by WIMPs, neutrons and neutrinos, from background-like electronic recoils (ER), generated by $\beta$- and $\gamma$-radioactivity and by neutrinos interacting with atomic electrons. The high density of LXe provides efficient self-shielding, allowing us to define a low-background, central detector region (fiducialization).  Dual-phase TPCs are leading the field in terms of sensitivity since several years~\cite{ref::xe10, ref::zeplin3, ref::xe100run10, ref::luxresult, ref::pandaX}. No dark matter signal has been observed yet~\cite{ref::dm2014}, excluding spin-independent WIMP-nucleon scattering cross sections above $7.6 \times 10^{-46}$\,cm$^2$ at a WIMP mass of 33\,GeV/$c^2$~\cite{ref::luxresult}.

The next generation of LXe experiments is currently in the design phase or under construction. With target masses beyond the ton-scale, they aim at being sensitive to cross sections down to a few $\times$10$^{-47}$\,cm$^2$ (XENON1T~\cite{ref::xe1t}, with exposures around 2\,\ex) or even to a few $\times$10$^{-48}$\,cm$^2$ (XENONnT~\cite{ref::xe1t_mv}, LZ~\cite{ref::lz}, 20\,\ex). This poses the question which sensitivities could be achieved by the next-to-next generation of instruments, which aim for multi-ton target masses and are for example studied within the DARWIN project~\cite{ref::darwin1, ref::darwin2, ref::darwinweb}. Such detectors will be eventually limited by irreducible background events from coherent neutrino-nucleus scattering (CNNS), which are a priori indistinguishable from WIMPs. At the same time, all other background sources must be kept under control in order to reach the best possible sensitivity. The low-background environment will open up a number of additional physics channels, among them the precise measurement of the low energy solar pp-neutrino flux~\cite{ref::darwinnu, ref::suzuki}, supernova neutrinos~\cite{ref::sn_neutrinos}, axions and axion-like particles~\cite{ref::axions,ref::darwinaxions}, as well as rare nuclear processes such as the neutrinoless double beta decays of $^{136}$Xe~\cite{ref::darwinnu} and $^{134}$Xe, $^{126}$Xe and $^{124}$Xe~\cite{ref::darwin0nbb}. However, the search for WIMP dark matter remains the prime science case for a DARWIN-like detector. It will be probed via spin-independent and spin-dependent WIMP-nucleon interactions, as $^\textnormal{\footnotesize nat}$Xe contains about 50\% isotopes with non-zero nuclear spin, and via inelastic processes~\cite{ref::inelastic}.

In this work we present a detailed study on the sensitivity reach of a DARWIN-type multi-ton LXe detector. We mainly focus on spin-independent interactions and take into account backgrounds, background rejection efficiency, exposure, thresholds and energy resolution. To this end, we determine the sensitivity, which we define as the average 90\% confidence level (CL) exclusion limit, for various combinations of these parameters by simulating the outcome of a large number of trial experiments. The analysis is performed by modeling all expected signals and backgrounds within a single framework. This allows us to formulate a number of requirements which must be considered in the design of such an experiment in order to explore cross sections as low as a few $\times$10$^{-49}$\,cm$^2$.

The article is structured as follows: the S1 and S2 signal generation in the simulation is described in Section~\ref{sec::generation}, while Section~\ref{sec::escale} explains how we use these signals to construct different energy scales, based on the combined S1+S2 signal or the S1 signal alone, with their impact on the energy resolution. Section~\ref{sec::limit} describes the method to evaluate the sensitivity to spin-independent WIMP-nucleon interactions, based on repeated toy-experiments with random background realizations. The background conditions are defined in Section~\ref{sec::background} and the required ER rejection level is discussed in Section~\ref{sec::rejection}. The main results of this study, the dependence of the WIMP sensitivity on various parameters, are presented in Section~\ref{sec::results}. The article closes with a summary and a discussion of our main results, including an extension to spin-dependent interactions.

\section{Light and Charge Signal Generation} 
\label{sec::generation}

In order to simulate the detection process of signal and background events in a realistic fashion, we employ a  signal generation model following~Ref.~\cite{ref::nest}. Based on the energy and the type of interaction, ERs measured in keVee (electronic recoil equivalent energy) or NRs measured in keVnr (nuclear recoil equivalent), the mean number of generated quanta $\overline{n}_\gamma$ (scintillation photons) and $\overline{n}_e$ (ionization electrons) is calculated for a drift field of 500\,V/cm, a typical number for dual-phase TPCs. For ERs, charge and light yields are taken from NEST v0.98~\cite{ref::nest}, while for NRs, $Q_y$ and ${\cal L}_\textnormal{\footnotesize eff}$ are taken from XENON100, using Refs.~\cite{ref::xe100ambe} and~\cite{ref::xe100run8}, respectively. $Q_y$ is only given down to 3\,keVnr and we extrapolated to lower energies by fixing it to the value at 3\,keVnr. The \leff  parametrization assumes  ${\cal L}_\textnormal{\footnotesize eff}=0$ for $E\le1$\,keVnr, with the consequence that no light is produced for energy depositions below this energy. The respective event is therefore discarded, as it would not be observed in the TPC. The total number of quanta $\overline{n} = \overline{n}_\gamma + \overline{n}_e$ is convoluted by a Gaussian distribution with a width of $\sigma = \sqrt{F \times \overline{n}}$, assuming a Fano factor $F=0.03$~\cite{ref::fano}, in order to get the generated number of quanta $q$. The number of scintillation photons, eventually leading to the light signal (S1), is randomly drawn from a Binomial distribution 
\begin{equation}
n^0_\gamma = \textnormal{Binomial}(q,p=\frac{\overline{n}_\gamma}{\overline{n}})\textnormal{.}
\end{equation}
The observable in the detector is the number of photoelectrons (PE)
\begin{equation}
n^0_{S1} = n^0_\gamma \times \epsilon
\end{equation}
with the photon detection efficiency 
\begin{equation}
\epsilon = \frac{L_y}{Y} \textnormal{.}
\end{equation}
$L_y$ is the average light yield in PE/keVee at 122\,keV and zero field and $Y = 63.4$ is the number of photons/keV generated in LXe at zero field~\cite{ref::nest}. Our benchmark value is $L_y =8$\,PE/keV, about $2\times$ larger than achieved in XENON100~\cite{ref::xe100} and similar to the LUX value~\cite{ref::lux}. The number of electrons from the charge signal (S2) is consequently
\begin{equation} 
n^0_{S2} = q - n^0_\gamma \textnormal{.}
\end{equation}
$n^0_{S1}$ and $n^0_{S2}$ are anti-correlated, and the resolution due to the statistical nature of the signal-generation process is properly modeled. The signal quanta observed in the detector, $n_{S1}$ and $n_{S2}$, are drawn from a Poisson or a Gaussian distribution, respectively, with the means $n^0_{S1}$ and $n^0_{S2}$ and resolutions tuned to match existing experimental data. This takes into account additional effects from the photoelectron detection process (``single photoelectron resolution'') as well as the conversion of ionization electrons to photoelectrons (Gaussian process with a mean of 20-30\,PE/e$^-$~\cite{ref::xe100singlee,ref::zepsinglee}). Throughout this study, the charge signal (S2) is given in electrons, and we do not consider charge losses during the drift of the electrons through the LXe, which requires an effective gas cleaning system and 100\% charge extraction into the gas phase (as achieved by XENON100~\cite{ref::xe100singlee}).

\section{Energy Scale and Energy Resolution}
\label{sec::escale}

\begin{figure}[b]
\centering 
\includegraphics[width=.495\textwidth]{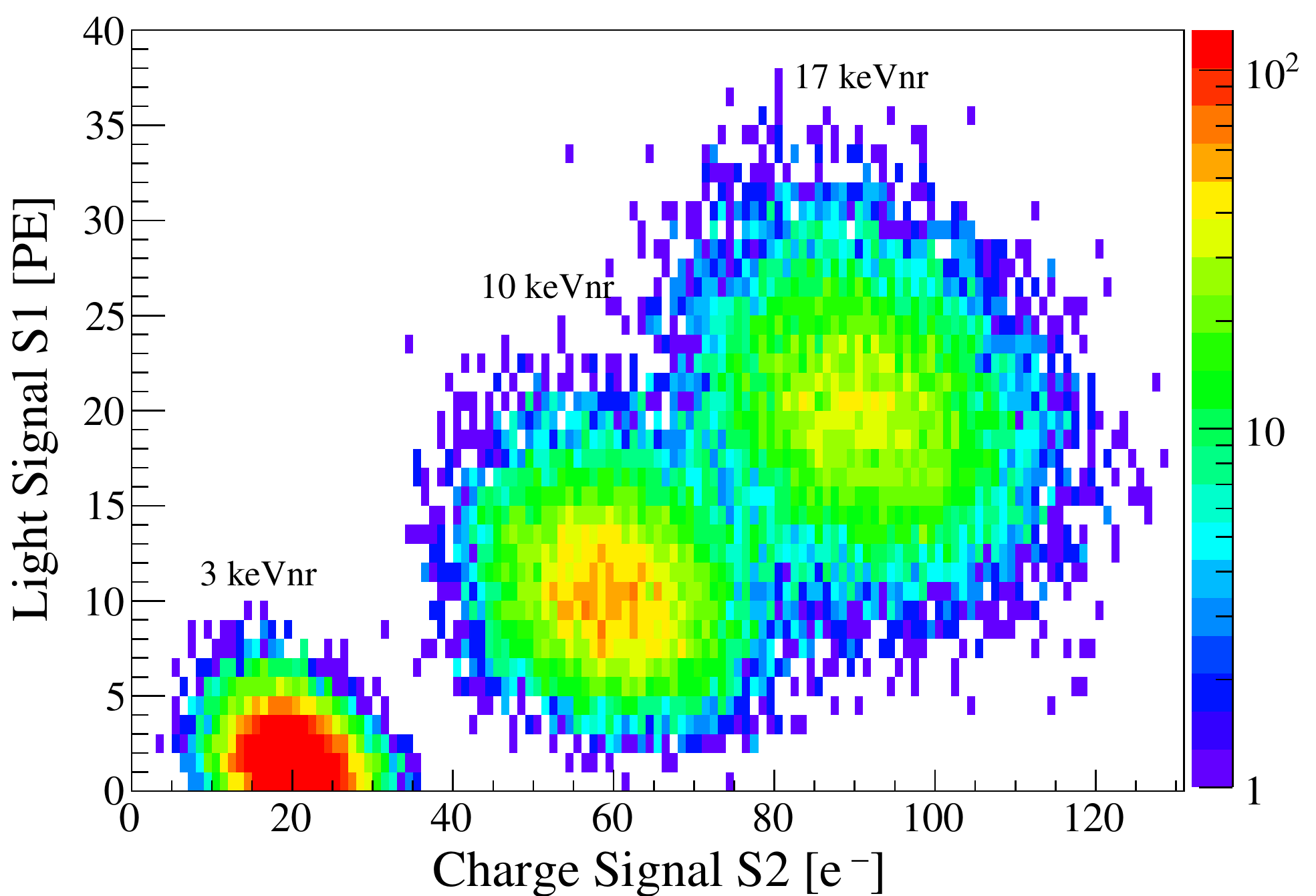}
\hfill
\includegraphics[width=.495\textwidth]{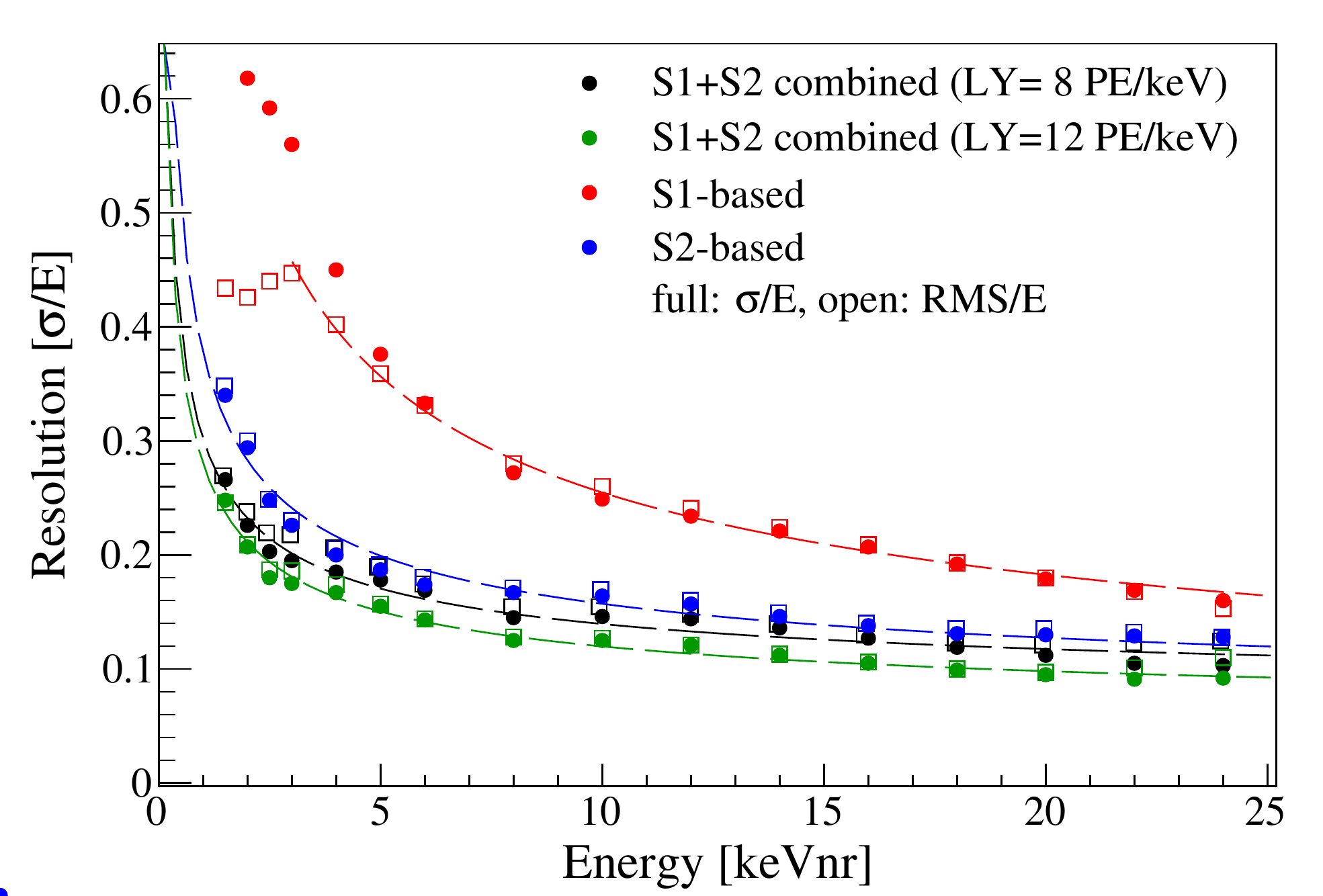}
\caption{{\bf (left)} Simulated light (S1) and charge signals (S2) from hypothetical mono-energetic~NR energy depositions. The same number of events was simulated for all three energies. A light yield $L_y=8.0$\,PE/keVee, full charge extraction, and the processes discussed in Section~\ref{sec::generation} are considered. The anti-correlation between both signals is very weak. {\bf (right)} Resolution of different energy scales as derived from the simulation. The ``combined energy'' scales use the S1 and S2 signals and take into account the anti-correlation. The data points are well described by functions of the form $a_0+a_1/\sqrt{E}$. As the light signal is non-symmetric at low energies, the RMS is a better measure for the resolution than the Gaussian width $\sigma$. \label{fig::escale} }
\end{figure}

In this work we consider two energy scales. The first one relies on the scintillation signal (S1) only, which is still the standard in the field and used in most published analyses~\cite{ref::xe100run10, ref::pandaX, ref::zeplin3}. The second one is based on light and charge (S1+S2), additionally taking into account the anti-correlation between the two observables, which is induced by the signal generation process: ionization electrons which recombine with Xe$^+_2$ molecules will be lost to the charge signal but eventually create scintillation light. This anti-correlation, though well known for high-energetic $\gamma$-lines~\cite{ref::anti}, has not yet been observed directly for low-energetic nuclear recoils. The event generation model described above allows us to establish the combined energy scale by simulating the light and charge signals generated by discrete NR energy depositions in the LXe, see Figure~\ref{fig::escale} (left). The anti-correlation is only very weak and hardly visible when the signal resolution is applied correctly. The energy scales are constructed based on the mean observed signals obtained for many mono-energetic energy depositions. Using the combined energy scale, which additionaly employs an anti-correlation angle, leads to a significant improvement in the energy resolution compared to the S1-only scale. However, the improvement compared to a charge-only scale is small, see Figure~\ref{fig::escale} (right). It is interesting to note that a $L_y$ increase from 8\,PE/keVee to 12\,PE/keVee does not lead to a significant improvement in the resolution of the combined scale.

We will see in the following that the CNNS background induced by solar $^8$B-neutrinos increases exponentially at very low NR energies. A good energy resolution is therefore mandatory in order to decrease the energy threshold for the WIMP search to values as close as possible to the rise. For an S1-based scale (8\,PE/keVee), one can achieve 6\,keVnr, while a threshold of 5\,keVnr or lower is only reachable when a combined energy scale is applied.

\section{Evaluation of the spin-independent WIMP-nucleon Sensitivity}
\label{sec::limit}

We evaluate the sensitivity of a DARWIN-type multi-ton LXe detector to spin-independent WIMP-nucleon scattering cross-sections by averaging the outcome of many simulated pseudo-experiments. The analysis is performed under realistic assumptions, i.e., the statistical nature of the signal generation (energy resolution) and of the observed background conditions is taken into account. All energy depositions in the detector, from signal and background, are converted into a {\it reconstructed NR energy}. For ER background events of ER energy $E_{ee}$, the respective S1 and S2 signals are generated according to Section~\ref{sec::generation}. From these observables, the reconstructed NR energy is calculated. For NR background events of energy $E_{nr}$, S1 and S2 signals are generated as well and also back-converted into reconstructed NR energy.

For each pseudo-experiment, the (integer) number of background events and their energies in reconstructed NR energy are determined by independently considering each of the background sources $X$ introduced in Section~\ref{sec::background} below. For an expected mean number of background events $b_X$, for a given exposure, energy interval and ER rejection level, the actual number of events $n_X$ is randomly drawn from a Poisson distribution with mean $b_X$. The energies of these events are distributed according to the energy spectra defined in Section~\ref{sec::background}, taking into account the energy scale under study. This means that background events from outside the WIMP analysis window, defined in reconstructed NR energy, can enter the region of interest (ROI) and events with initial energies in the window can be lost if the reconstructed NR energy is not in the ROI. The total number of background events is $n_b$. 
  
The expected spectra for WIMPs interacting in a LXe target of total mass $M$ are calculated from the differential event rate
\begin{equation}\label{eq::rate}
\frac{dR}{dE_{nr}} = \frac{\rho_0 \ M}{m_{Xe} \ m_\chi} \int_{v_{min}}^{\infty} v f(v) \frac{d \sigma}{dE_{nr}} \ dv\textnormal{.}
\end{equation}
$m_{Xe}$ and $m_\chi$ are the masses of the xenon nucleus and the WIMP, respectively, and $f(v)$ is the normalized WIMP velocity distribution. We take into account the masses and abundances of the different xenon isotopes occurring in natural xenon gas. All velocities are defined in the detector's reference frame, with \begin{equation}\label{eq::mu}
v_{min}=\sqrt{\frac{E_{nr} m_{Xe}}{2} \frac{(m_{Xe}+m_\chi)^2}{(m_{Xe} m_\chi)^2}} = \sqrt{\frac{E_{nr} m_{Xe}}{2} \frac{1}{\mu^2}}                                                                                                                                                                                                                        
\end{equation}
being the minimal velocity required to induce a nuclear recoil $E_{nr}$. The velocity distribution $f(v)$ is truncated at the escape velocity $v_{esc}=544$\,km\,s$^{-1}$~\cite{ref::rave}, the maximum velocity of WIMPs bound in the potential well of the galaxy. The canonical value for the local WIMP density at the Sun's position in the Galaxy is $\rho_{0}=0.3$\,GeV\,$c^{-2}$\,cm$^{-3}$. Because of its large de~Broglie wavelength, the WIMP interacts coherently with all nucleons in the target nucleus. We mainly focus on spin-independent (SI)  interactions for which the WIMP-nucleus scattering cross section in Eq.~(\ref{eq::rate}) is given by
\begin{equation}
\frac{d\sigma}{dE_{nr}} = \frac{m_N}{2 v^2 \mu^2}  \sigma_{SI} F^2(E_{nr}) \textnormal{.}
\end{equation}
The loss of coherence for heavy WIMP targets such as xenon is accounted for by the finite form factor $F(E_{nr})$, for which we use Helm's definition. The cross section reads
\begin{equation}\label{eq::si}
\sigma_{SI}=\sigma_n \frac{\mu^2}{\mu_n^2}\frac{(f_pZ+f_n(A-Z))^2}{f_n^2} = \sigma_n \frac{\mu^2}{\mu_n^2} A^2 \textnormal{,}
\end{equation}
where the $f_{p,n}$ describe the WIMP couplings to protons and neutrons. The second equality assumes $f_p=f_n$, leading to an $A^2$ dependence of the cross section. $\mu$ is the WIMP-nucleus reduced mass, see Eq.~(\ref{eq::mu}), and $\mu_n$ the one of the WIMP-nucleon system. It is used to relate the WIMP-nucleus cross section $\sigma$ to the WIMP-nucleon cross section $\sigma_n$, which is the relevant parameter for this study and allows the comparison with other target nuclei.

For a single event, the reconstructed NR energy $E'_{nr}$ is obtained by generating the corresponding S1 and S2 signals and the subsequent back-conversion using the previously defined energy scale. The observed number of events in a simulated toy experiment running for a live-time $T$ is obtained by integrating Eq.~(\ref{eq::rate}) from the threshold energy $E'_{low}$ to the upper boundary  $E'_{high}$:
\begin{equation}
N=T \int_{E'_{low}}^{E'_{high}} dE'_r \ \epsilon(E'_{nr}) \ \frac{dR}{dE'_{nr}} \textnormal{.} 
\end{equation}
In most parts of this study, we take an energy-independent efficiency $\epsilon=0.3$ for NRs. As $dR/dE_{nr}$ is a steeply falling exponential function, $E'_{high}$ is much less relevant than the energy threshold $E'_{low}$ and fixed to 20.5\,keVnr or 35.0\,keVnr. 

The WIMP sensitivity is evaluated in two ways: in order to study how the sensitivity depends on various parameters we perform an analysis based on a fixed WIMP search window, which is defined from $E'_{low}$ to 20.5\,keVnr and by a fixed ER rejection level at 30\% NR acceptance. For each simulated trial, all background events in the box are considered as possible WIMP signals, and the 90\% CL exclusion limit is calculated according the maximum gap method~\cite{ref::yellin}. By construction, this method takes into account the different spectral shapes of signal and background. The sensitivity is determined by averaging the results of 1000~individual simulations. The advantage of this approach is that it is straight-forward to quantify the number of signal and background events in the region of interest, and allows for an easy variation of detector parameters. The upper energy limit of 20.5\,keVnr was chosen as the signal-to-background ratio starts to deteriorate at higher energies, especially for small ($\sim$10\,GeV/$c^2$) and medium WIMP masses ($\sim$100\,GeV/$c^2$), and the gain in sensitivity is small.

Systematic uncertainties, such as the finite knowledge on the energy scale which should be considered in a proper 90\% CL exclusion limit, are beyond the scope of this study and neglected. It is known that likelihood methods, also taking into account the event distribution in discrimination space $D$\,$\propto$\,S2/S1, can improve the sensitivity by a factor $\sim$2~\cite{ref::xe100_pl} and additionally include systematic uncertainties in a proper way. For the optimal parameters found by the study with the fixed WIMP search window, we calculate the sensitivity using a likelihood method. The natural logarithm of the likelihood is defined as
\begin{eqnarray} \label{eq::ll}
\log {\cal L} = - (\mu_\chi + b_{ER} + b_{NR}) + \sum_{k=1}^{n_{tot}} \log &[& \mu_\chi f_\chi(E'_{nr,k}) \ g_\chi(D_k)  + b_{ER} \ f_{ER}(E'_{nr,k}) \ g_{ER}(D_k) \nonumber \\ &+& b_{NR} \ f_{NR}(E'_{nr,k}) \ g_{NR}(D_k) \ ],
\end{eqnarray}
where $\mu_\chi$ and $b_X$ are the expected mean values for WIMPs and ER and NR backgrounds. $f_X(E'_{nr,k})$ and $g_X(D_k)$ denote the normalized probability distribution functions of event source~$X$ in energy $E'_{nr}$ and discrimination parameter space $D$, respectively. The sum is evaluated for all for all $n_{tot}$ events generated in each toy experiment. The test statistic is defined as 
\begin{equation}
q = \begin{cases}-2 \log\left( \frac{{\cal L}(\mu_\chi)}{{\cal L}(\hat{\mu}_\chi)} \right) = 2 \log {\cal L}(\hat{\mu}_\chi) - 2 \log {\cal L}(\mu_\chi) &\quad \textnormal{for} \ \hat{\mu}_\chi \leq \mu_\chi \\ 0 &\quad \textnormal{for} \ \hat{\mu}_\chi > \mu_\chi, 
 \end{cases}
\end{equation}
where $\hat{\mu}_\chi$ is Maximum Likelihood estimator of Eq.~(\ref{eq::ll}). Its distributions under the signal $H_{\mu_\chi}$ and background hypothesis $H_0$ are used to derive the mean sensitivity. The procedure takes into account the CLs mechanism to protect the result against statistical downward fluctuations of the background~\cite{ref::cls}. We generally follow~\cite{ref::cowan,ref::xe100_pl} where more details can be found.

\section{Backgrounds}
\label{sec::background}

We consider all relevant backgrounds for multi-ton scale LXe dark matter detectors, which can be separated into two categories. External backgrounds comprise $\gamma$-rays and neutrons, which stem from radioactive decays or interactions outside of the LXe target, for example in the detector's construction materials. Due to the high density of LXe ($\rho \approx 3$\,g/cm$^3$), they can be reduced considerably by target fiducialization. This does not work for intrinsic backgrounds, which are uniformly distributed in the target region.

In the following, we detail the background assumptions which enter our analysis. All contributions leading to single scatter signatures in the low-energy WIMP search region are summarized in Table~\ref{tab::background}. If the background distribution is not flat, we quote the average values for a given energy interval. All numbers are given assuming an infinite energy resolution while for the sensitivity study presented below, the resolution is taken into account.

\begin{table}[]
\centering
\begin{tabular}{|l|cll|}
\hline
Source & Rate  & Spectrum & Comment\\
 & \footnotesize{[events/(t$\cdot$y$\cdot$keVxx)]} & & \\  
\hline
$\gamma$-rays materials & 0.054\textcolor{white}{000.} & flat & assumptions as discussed in text\\
neutrons$^*$ & 3.8$\times$10$^{-5}$ & exp.~decrease & average of [5.0-20.5]\,keVnr interval\\
intrinsic $^{85}$Kr & 1.44\textcolor{white}{0000.} & flat & assume 0.1\,ppt of $^\textnormal{\footnotesize nat}$Kr\\
intrinsic $^{222}$Rn & 0.35\textcolor{white}{0000.} & flat & assume 0.1\,$\mu$Bq/kg of $^{222}$Rn\\
$2\nu\beta\beta$ of $^{136}$Xe & 0.73\textcolor{white}{0000.} & linear rise & average of [2-10]\,keVee interval \\
pp- and $^7$Be $\nu$ & 3.25\textcolor{white}{0000.} & flat & details see~\cite{ref::darwinnu} \\
CNNS$^*$ & 0.0022\textcolor{white}{00.} & real & average of [4.0-20.5]\,keVnr interval\\
\hline
\end{tabular}
\caption{\label{tab::background} Background contributions considered in the sensitivity analysis. All values given here are before ER discrimination or a finite NR acceptance and do not yet take into account an energy resolution. The ER background rates are given per keVee (electron recoil equivalent). The NR backgrounds, marked by the asterisk, are given per keVnr (nuclear recoil equivalent). If the background is not flat, an average value over a finite interval is quoted. The energy resolution is taken into account for the sensitivity study. This is of particular importance in case of the steeply rising CNNS spectrum. }
\end{table}

\paragraph{$\gamma$-rays from materials}

The $\gamma$-background produced in the laboratory is virtually zero behind a shield of several meters of water~\cite{ref::xe1t_mv}. The remaining background stems from radioactive contamination (mainly from the $^{238}$U, $^{235}$U and $^{232}$Th chains, $^{40}$K, $^{60}$Co) in the cryostat and detector materials. It can be reduced by target fiducialization and discrimination using the charge-to-light (S2/S1) ratio. Based on earlier  studies~\cite{ref::darwinnu}, we assume a flat background spectrum and a rate of 0.054\,events/(t$\cdot$y$\cdot$keVee) before discrimination, which is reached behind a LXe layer of $\sim$13\,cm. $\alpha$- and $\beta$-backgrounds from detector materials are irrelevant even for small LXe TPCs as they do not penetrate into the fiducial target region.

\paragraph{Neutrons}

As WIMPs are expected to generate nuclear recoils in the detector, background from neutrons, producing the same signature, is dangerous as the S2/S1-background rejection cannot be used. However, one can rely on target fiducialization (even though neutrons have a considerably longer mean free path than $\gamma$-rays of the same energy, and have also higher energies) and on rejection based on the event multiplicity. Background neutrons are either of cosmogenic origin, i.e., produced in muon induced electromagnetic and hadronic showers, or radiogenic, i.e., generated in $(\alpha,n)$ or spontaneous fission of heavy isotopes. A shielding of several meters of water, operated as a \u{C}erenkov muon veto, decreases the muon-induced neutron background to very low levels~\cite{ref::xe1t_mv}, and radiogenic neutrons from the laboratory are efficiently reduced by many orders of magnitude. Both backgrounds are therefore considered negligible in this study.

More relevant are radiogenic neutrons emitted from the detector materials itself. The light PTFE, usually used as insulator and light reflector~\cite{ref::ptfe_reflect}, is a known source of ($\alpha,n)$ neutrons, where the $\alpha$-particle is from $^{238}$U, $^{235}$U and $^{232}$Th-chain contaminations. Another critical component are the photosensors which are made from a variety of materials, sometimes without low-background alternatives~\cite{ref::xe1tpmt}. Here we assume a NR background from radiogenic neutrons which is dominated by the photosensors and contributes a rate of $3.8 \times 10^{-5}$\,events/(t$\cdot$y$\cdot$keVnr) (average value in a 5.0-20.5\,keVnr interval). This is comparable to the ER background from two-neutrino double-beta decays ($2\nu\beta\beta$) when taking into account realistic ER rejection and NR acceptance levels. This neutron background rate is motivated by a Monte Carlo simulation performed for an an idealized DARWIN-type detector with a 18\,t LXe target, where the relevant components are assumed to be made from OFHC (cryostat vessels, electrodes, diving bell), PTFE (TPC) and quartz (photosensors). We have optimized the detector model presented in Ref.~\cite{ref::darwinnu} for a low neutron background by reducing the TPC PTFE reflectors to a thickness of 5\,mm, and by assuming cleaner materials (PTFE factor~4, e.g., entry~43 in~\cite{ref::exo}; copper factor~2, e.g., the ``cryostat copper'' in~\cite{ref::gator}, and photosensors factor~5). These would have to be identified in dedicated material screening and selection campaigns. The contributions of the photosensors could possibly be reduced by using improved low-background sapphire instead of quartz~\cite{ref::xe1tpmt} or by employing novel sensors such as gaseous photomultipliers (GPMs)~\cite{ref::gpm}.

\begin{figure}[h!]
\centering 
\includegraphics[width=.495\textwidth]{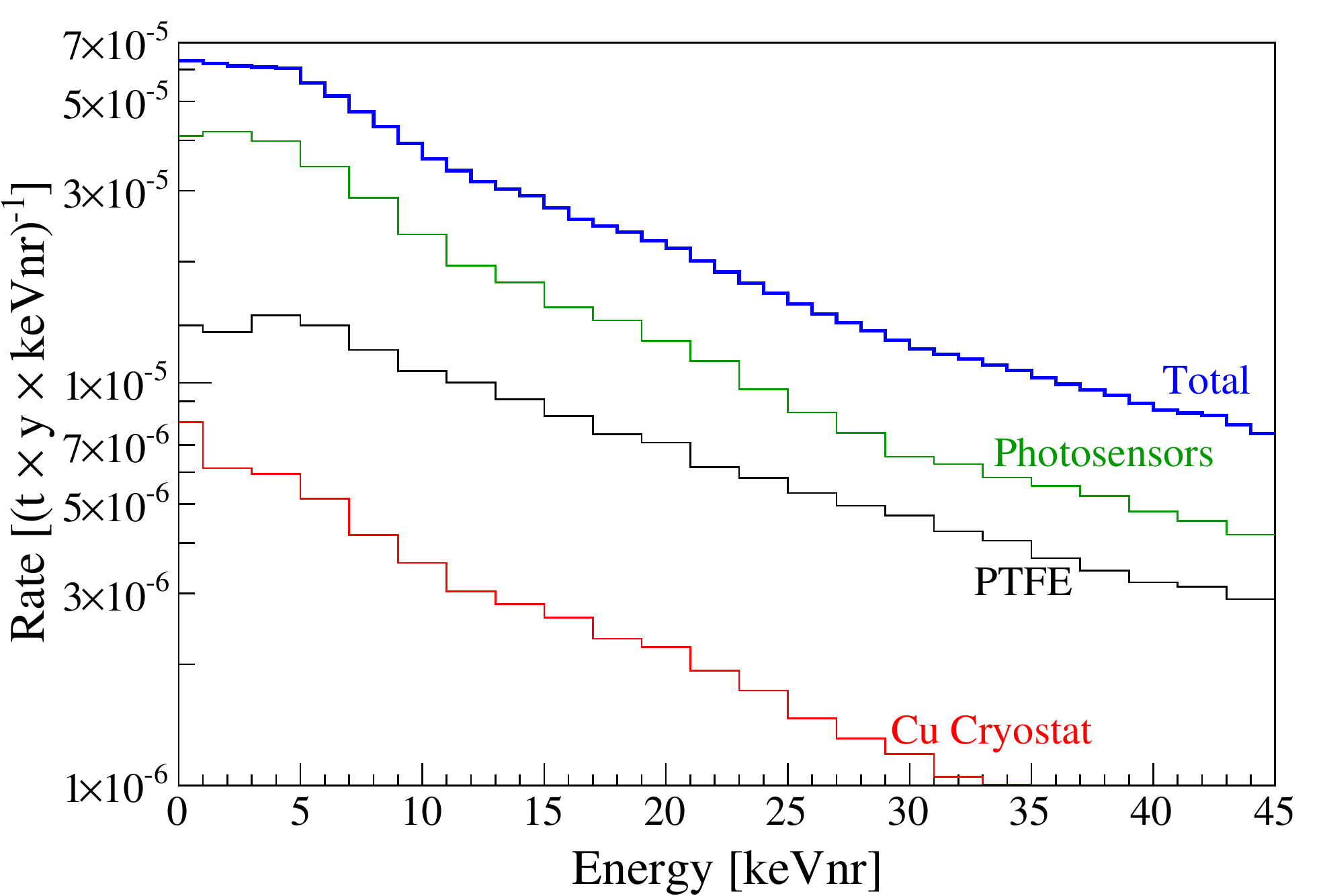}
\hfill
\includegraphics[width=.495\textwidth]{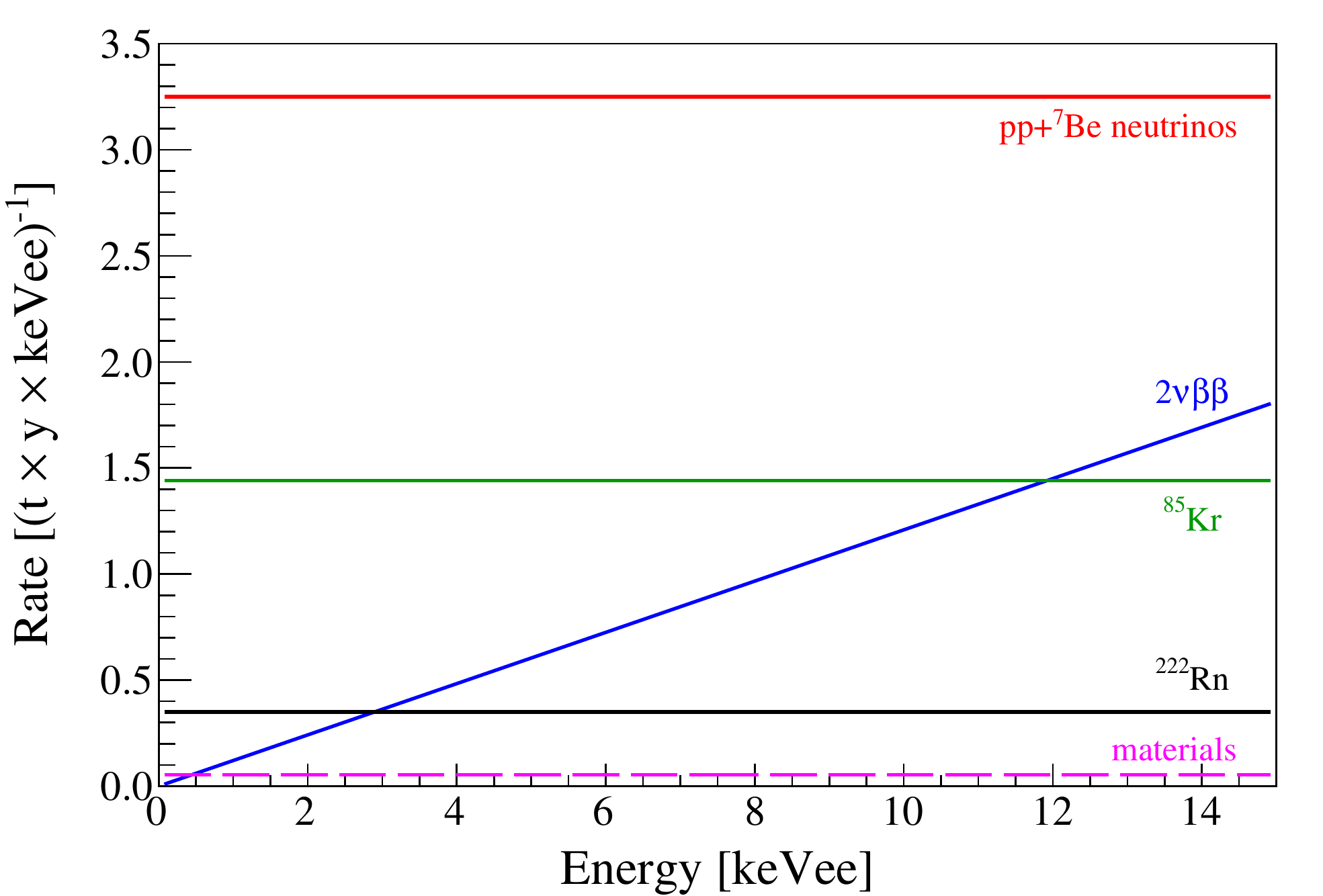}
\caption{{\bf (left)} Differential energy spectrum of single-scatter NR background from the detector materials, assuming 100\% NR acceptance and an infinite energy resolution.  {\bf (right)} Differential energy spectra of the ER background sources. No energy resolution or ER rejection is applied at this point. The dominating contribution is from pp- and $^7$Be solar neutrinos.  \label{fig::background} }
\end{figure}

The neutron production rates and energy spectra were calculated using the SOURCES-4A code~\cite{ref::sources}, based on the assumed radioactive contamination of the materials. Details on the calculations can be found in~\cite{ref::xe100_nr}. For each component (the two cryostats and the two photosensor arrays were simulated independently), $4\times 10^7$ neutrons were simulated in Geant4~\cite{ref::geant4}, which leads to a negligible statistical uncertainty. The finite efficiency to identify multiple scatter signatures increases the true single scatter background by a factor~1.85 when taking into account a resolution of 3\,mm (10\,mm) to separate vertices in the $z$ ($xy$) coordinate and assuming a realistic lower threshold to identify the second S2~signal in double scatters~\cite{ref::xe100_nr}. Figure~\ref{fig::background} (left) shows the exponentially falling recoil spectrum, which is reached after a fiducial cut of $\sim$16\,cm from all sides. As the surface-to-volume ratio decreases with increasing detector size, the background situation will improve for detectors larger than the simulated one, as the background sources scale with the surface. For this study, we do not pick a specific geometry but assume that the neutron background scales with exposure, similar to target-intrinsic backgrounds. We note that a full study of this background contribution requires the precise knowledge of the detector components for the neutron production (contamination, material assembly), as well as detector size and geometry, which is not defined at this point.

\paragraph{Intrinsic $^{85}$Kr}

Xenon does not have long-lived isotopes besides $^{136}$Xe, a $2\nu\beta\beta$ emitter, and can be purified to a high degree from other contaminations. One exception are other noble gases, as these are chemically inert (cannot by removed by chemical methods) and mix very well into xenon. As a consequence, such backgrounds are uniformly distributed in the target. A problematic isotope is the anthropogenic $\beta$-emitter $^{85}$Kr ($T_{1/2} = 10.76$\,y), present in natural krypton at the $2 \times 10^{-11}$\,g/g~level~\cite{ref::krconc}.
Krypton can be separated from xenon by cryogenic distillation, exploiting its 10$\times$ higher vapor pressure at LXe temperatures~\cite{ref::xmasscolumn}, and $^\textnormal{\footnotesize nat}$Kr-concentrations of less than 1\,ppt in xenon have already been achieved~\cite{ref::lindemann}. At low energies, the single scatter ER spectrum from $^{85}$Kr is basically flat~\cite{ref::xe100_er}. For this study we assume a $^\textnormal{\footnotesize nat}$Kr concentration of 0.1\,ppt, only a factor of $\sim$5~below the design goal of systems currently under development for the next generation projects~\cite{ref::krcolumn}. It leads to 1.44\,events/(t$\cdot$y$\cdot$keVee) before S2/S1 discrimination~\cite{ref::darwinnu}, see also Figure~\ref{fig::background} (right).

\paragraph{Intrinsic $^{222}$Rn}

The radioactive noble gas $^{222}$Rn is part of the $^{238}$U chain. It decays with a short half-life of 3.8\,d, however, it is constantly re-emanated from surfaces due to the presence of $^{226}$Ra. Therefore, all materials in contact with xenon anywhere in the detector system must be selected for their low Rn-emanation~\cite{ref::rnemanat}. The DARWIN background needs to be around 0.1\,$\mu$Bq  per kg of LXe, about an order to magnitude below the goal of the next generation experiments. The resulting single scatter rate before discrimination is 0.35\,events/(t$\cdot$y$\cdot$keVee) with a flat recoil spectrum~\cite{ref::darwinnu}. While the improved surface-to-volume ratio is generally favorable for the reduction of this background in very large detectors, an active removal of $^{222}$Rn from the LXe target is difficult at these low concentrations and would require further R\&D. Adsorption on charcoal~\cite{ref::rnremoval}, for example, would need enormous amounts of additional xenon and charcoal with a sufficiently low $^{226}$Ra contamination.

\paragraph{Two-neutrino double-beta decay of $^{136}$Xe}

We assume that the DARWIN facility is initially operated with natural xenon, which contains 8.9\% of $^{136}$Xe known to decay via a two-neutrino double-beta process ($2\nu\beta\beta$) with a half-life of $T_{1/2}= 2.17 \times 10^{21}$\,y~\cite{ref::exo2nbb}. Even though this is 11~orders of magnitude longer than the age of the Universe, its continuous single-scatter spectrum ($Q$-value is $(2458.7 \pm 0.6)$\,keV~\cite{ref::qval}) contributes to the WIMP search background. At lowest energies, its ER spectrum can be well approximated by a linear function starting from $(0,0)$. The integral rate in an 2-10\,keVee interval is 5.8\,events/(t$\cdot$y) before ER rejection~\cite{ref::darwinnu}. In principle this background can be completely avoided by using a target depleted of $^{136}$Xe.

\paragraph{Low-energy solar neutrinos}

The vast majority of neutrinos emitted from the Sun are neutrinos generated in the pp-fusion process or the subsequent $^7$Be reaction. Due to their rather low energies and high abundance, together with the impossibility to reduce their contribution by target purification, fiducialization or single-scatter selection, they are the most relevant source of ER background for LXe-based dark matter searches beyond the ton-scale. At lowest energies, they generate a basically flat ER spectrum and can only be separated from WIMP-induced NRs based on their S2/S1 ratio, albeit with a finite efficiency. The combined rate of pp- and $^7$Be neutrinos, which are indistinguishable in a DARWIN-like detector, is 3.25\,events/(t$\cdot$y$\cdot$keVee) before discrimination~\cite{ref::darwinnu}. Extending the energy range beyond the WIMP search region to $\sim$30\,keVee allows the precise measurement of the low-energy solar neutrino flux with DARWIN.

\begin{figure}[t]
\centering 
\includegraphics[width=.495\textwidth]{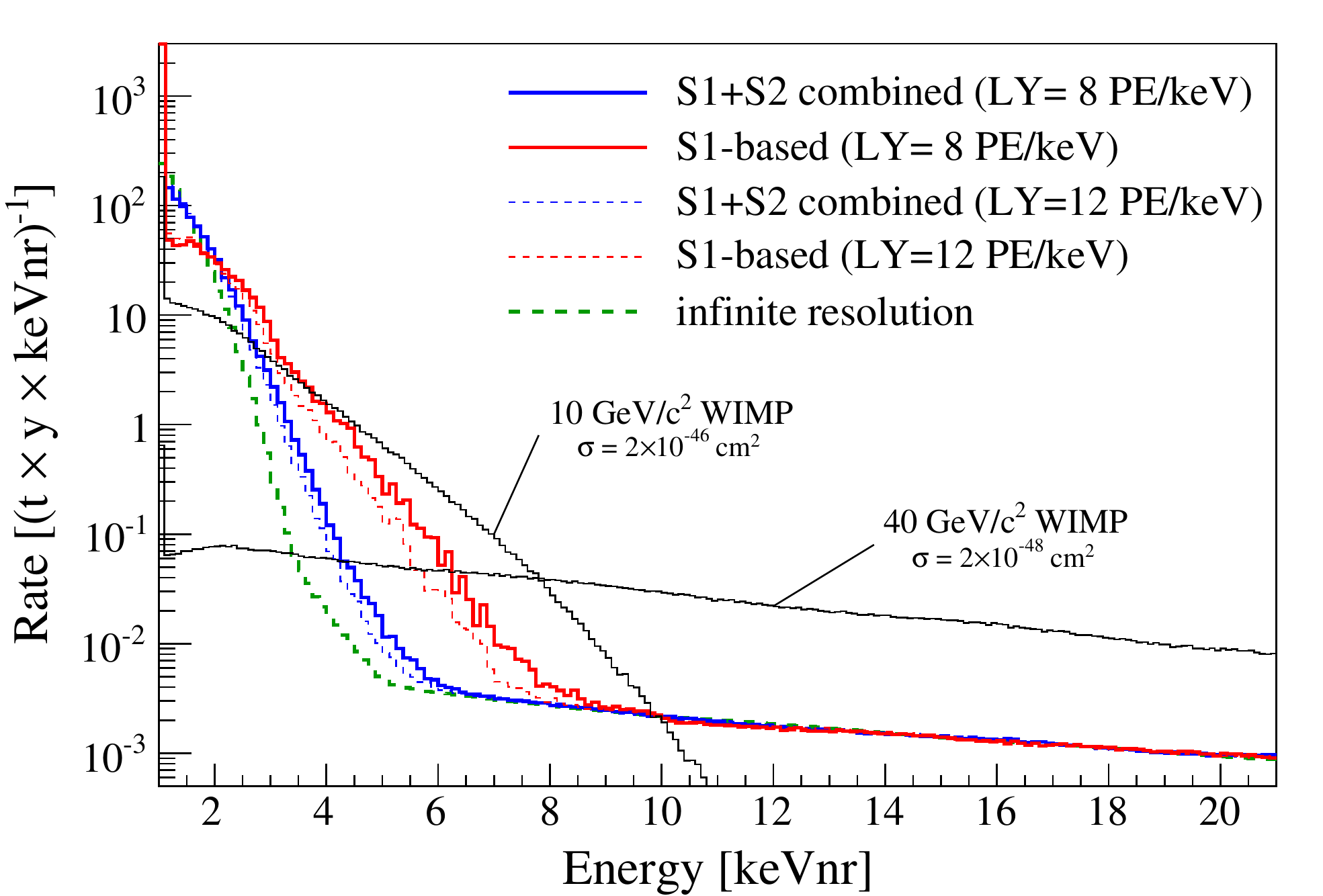}
\hfill
\includegraphics[width=.495\textwidth]{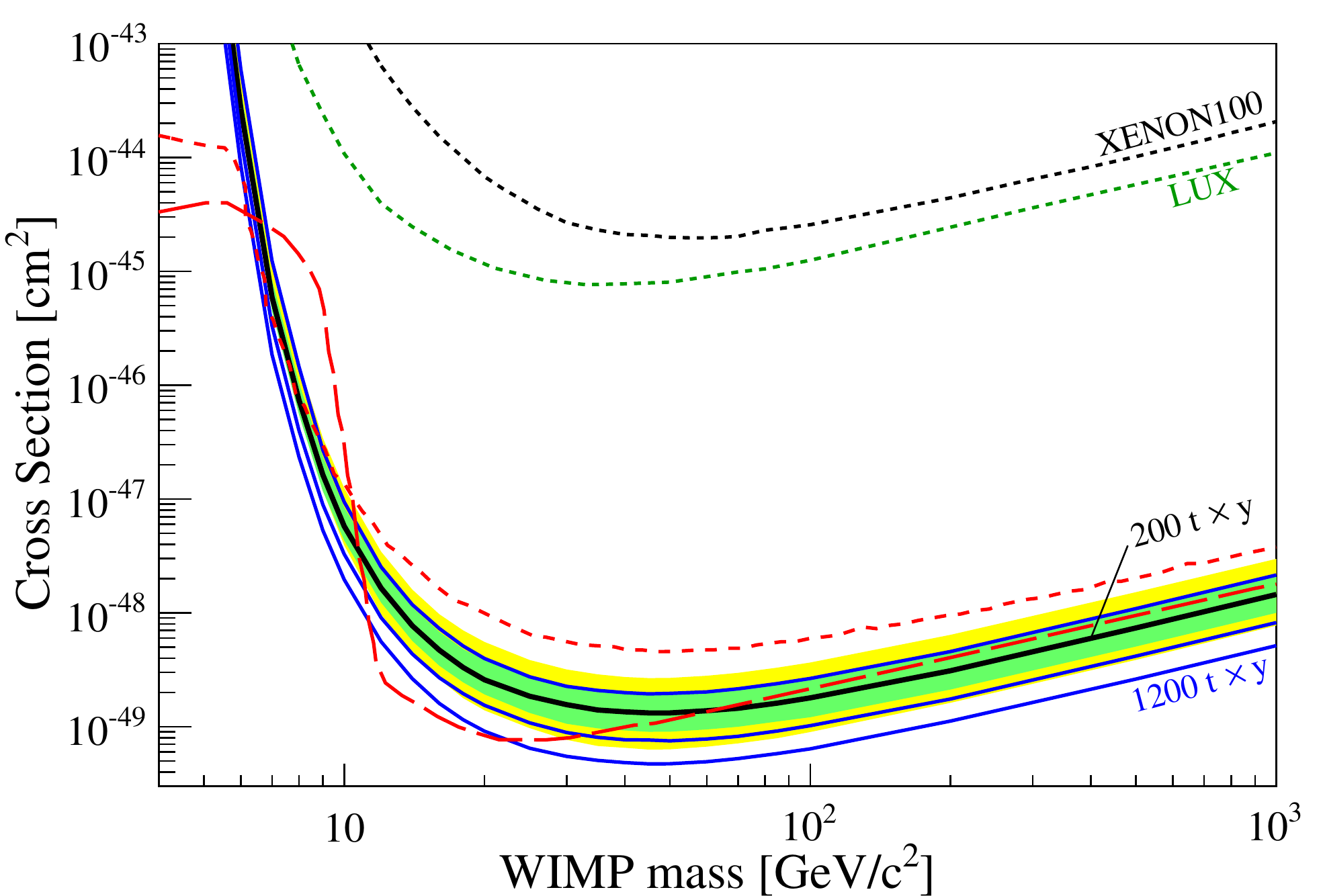}
\caption{{\bf (left)} Differential CNNS spectrum for different energy scales, based on S1-only or S1+S2. The largest improvement in energy resolution can be achieved by using light and charge signal; the impact of a light yield increase by 50\% is only mild. The spectrum with infinite resolution (dashed) is taken from~\cite{ref::strigari}. The steep rise below 4.0\,keVnr is from solar $^8$B neutrinos. We assume that no light is produced for $E_{nr}<1$\,keVnr, leading to the large number of entries in the lowest bin of the reconstructed spectra. Using the S1+S2~scale with 8\,PE/keVee, we also show the expected recoil spectra for WIMPs of 10\,GeV/$c^2$ and 40\,GeV/$c^2$ with cross sections of 2\,$\times$\,10$^{-46}$\,cm$^2$ and 2\,$\times$\,10$^{-48}$\,cm$^2$, respectively. {\bf (right)} WIMP exclusion limits at 90\% CL ignoring all backgrounds besides the one from CNNS for exposures of 100\,\ex, 200\,\ex, 500\,\ex \ and 1200\,\ex. For the likelihood analysis we use a S1+S2 combined energy scale with a light yield of 8\,PE/keVee at 122\,keVee, a 5-35\,keVnr energy interval and assume an unrealistic NR acceptance of 100\%. (The ER rejection will lower this value.) The result for 200\,\ex \ is shown with its 1$\sigma$ and 2$\sigma$ intervals
together with the published limits from XENON100~\cite{ref::xe100run10} and LUX~\cite{ref::lux}. For comparison the ``WIMP discovery limit'' (red dashed) and the ``1~event line'' (red dotted) of~\cite{ref::billard} are shown as well. 
\label{fig::cnns} }
\end{figure}

\paragraph{Coherent neutrino-nucleus scattering}

Dark matter WIMPs are expected to produce single-scatter NR events in the LXe detector. The identical signature is generated by neutrinos, which are predicted to coherently scatter off the xenon nucleus. Therefore, it is impossible to a priori distinguish the two sources of events. As coherent neutrino-nucleus scattering (CNNS) is a standard model process, and neutrinos of the right energies to produce signals in the WIMP search energy region exist, this background will ultimately limit the sensitivity of any WIMP detector~\cite{ref::billard}, if no additional information on the direction of the NR track is available~\cite{ref::directionality}. For a LXe target, the relevant neutrinos are mainly $^8$B neutrinos from the Sun, generating a ``wall'' of events rising steeply below $\sim$4\,keVnr, which significantly reduces the sensitivity to 5-8\,GeV/$c^2$ low-mass WIMPs. At higher NR energies, the main CNNS background is from atmospheric neutrinos, with smaller contributions from solar neutrinos from the helium-proton reaction (hep) and the diffuse supernova neutrino background (DSNB)~\cite{ref::strigari}. The background from CNNS strongly depends on the threshold, see Figure~\ref{fig::cnns} (left), which renders the energy scale (resolution) used for the analysis crucial. At higher NR energies, the rate is fairly flat around 1-2$\ \times 10^{-3}$\,events/(t$\cdot$y$\cdot$keVnr).

We note that the ``WIMP discovery limit'', as introduced in Ref.~\cite{ref::billard} for every WIMP mass, corresponds to the cross section at which a WIMP can be detected at 3$\sigma$ given a background of 500\,CNNS events above a LXe threshold of 4\,keVnr (no energy resolution applied). A utopic LXe exposure of 5300\,\ex \ is required to reach this number of background events in a 4-35\,keVnr interval, even assuming 100\% NR acceptance. Being somewhat more realistic, we calculate the average 90\% CL exclusion limit for exposures from 100\,\ex \ to 1200\,\ex, considering only the CNNS background assuming 100\% NR acceptance. We use the combined energy scale of Section~\ref{sec::escale} with a light yield of 8.0\,PE/keVee and an energy range of 5-35\,keVnr. The result is shown in Figure~\ref{fig::cnns} (right) and compared to the ``WIMP discovery limit'' and the ``1~event line'' of~\cite{ref::billard}.

\section{Electronic Recoil Rejection}
\label{sec::rejection}

In order to reach sensitivities which are limited by the ``ultimate'' NR background from CNNS, the ER backgrounds from intrinsic radioactive impurities, two-neutrino double-beta decay and especially solar neutrinos interacting with atomic electrons have to be reduced significantly. The lifetimes of the xenon singlet and triplet excimer states, which produce the scintillation light, are not very different, rendering pulse-shape discrimination inefficient~\cite{ref::psd}. Therefore, LXe-based dual-phase TPCs rely on signal-background discrimination based on the different charge-to-light ratio (S2/S1) for ERs and NRs, caused by the different energy loss mechanisms~\cite{ref::discrimination}: at a given energy deposition, ERs exhibit a larger S2~pulse than NRs. Rejection levels of 99.5\%~\cite{ref::xe100}, 99.6\%~\cite{ref::luxresult} and 99.987\%~\cite{ref::zeplin3} at 50\% NR acceptance have been already achieved in dark matter detectors, where the rejection appears to improve towards lower recoil energies. In this study, we assume flat, average values for rejection and acceptance. We also ignore possible pathological signatures which could lead to so-called ``anomalous'' leakage into the WIMP signal region. Such signatures could come from incomplete light or charge collection or from accidental coincidences of causally non-connected pulses. 

\begin{figure}[t]
\begin{minipage}[]{0.49\textwidth}
\begin{center}
\includegraphics[width=1.0\textwidth]{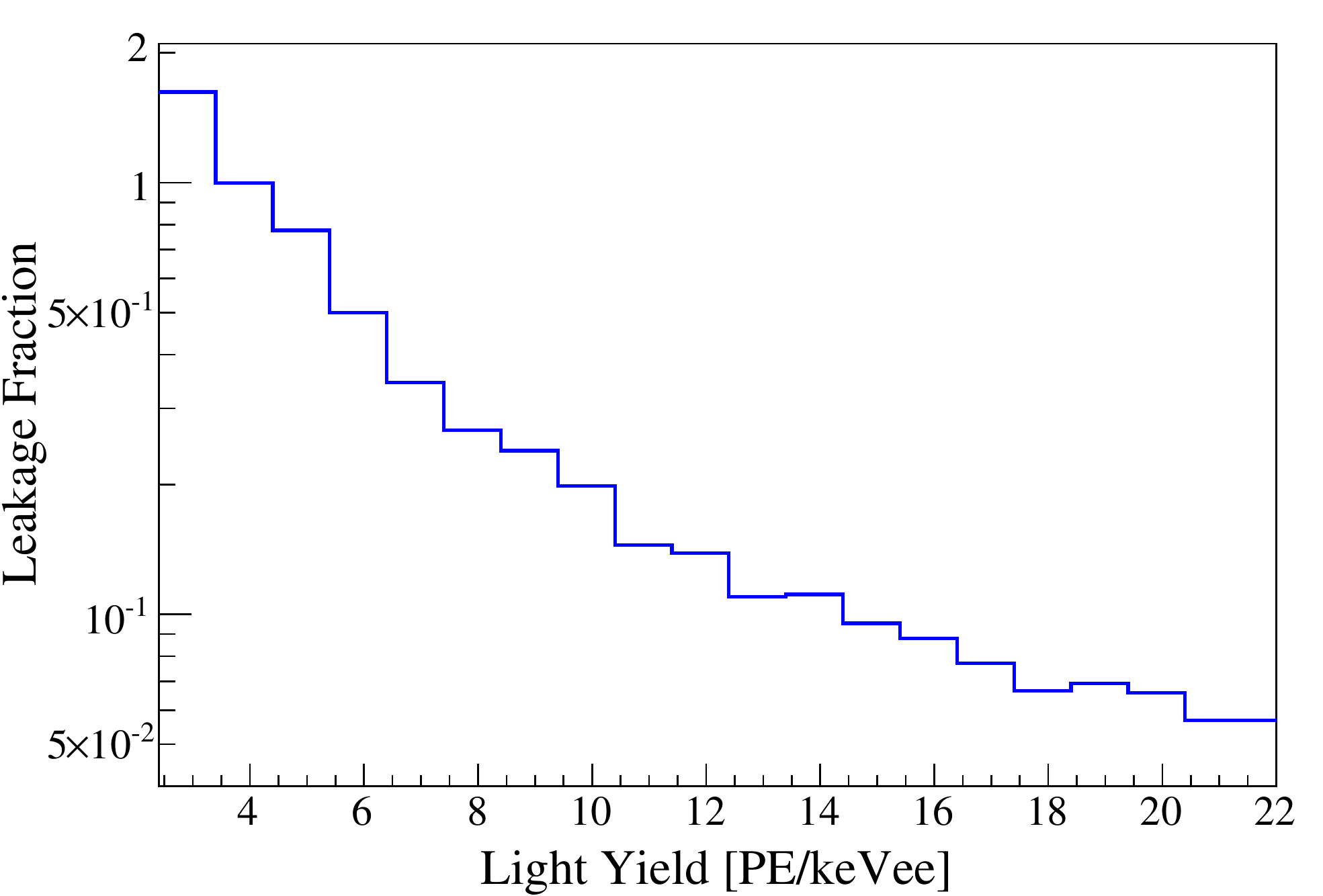}
\end{center}
\end{minipage}
\hfill
\begin{minipage}[]{0.49\textwidth}
\caption{Illustration of the impact of the zero-field light yield $L_y$ at 122\,keVee on the ER rejection, assuming a constant NR acceptance of 30\%. Using the signal model introduced in Section~\ref{sec::generation}, the fraction of ER events leaking into a low energy WIMP search region is determined in a simulation, and is quoted relative to the leakage for a moderate light yield of 4.0\,PE/keVee. All other parameters which might have an impact on the ER rejection were kept constant. The fluctuations in the leakage fraction are purely statistical. \label{fig::rejection_mc} }
\end{minipage}
\end{figure}

The mean values of the S2/S1 vs.~energy distributions for ERs and NRs are determined by the signal generation processes, which are affected by the drift field in the TPC: a lower drift field leads to more charge recombination and hence a smaller S2~but a larger S1~signal. The width of the distributions, crucial to achieve high ER rejection at high NR acceptance, is determined by the intrinsic fluctuations in the generation of the initial quanta and the statistical fluctuations in the detection processes. The latter can be mainly addressed by the design of a multi-ton scale detector, for example by maximizing the light and charge yields. By using the signal generation model introduced in Section~\ref{sec::generation}, we illustrate the impact of the light yield on the ER rejection in Figure~\ref{fig::rejection_mc}. We estimate the fraction of ER events leaking into a low energy WIMP search region, which is defined by a fixed 30\% NR acceptance. A moderate light yield of 4.0\,PE/keVee, as realized by XENON100~\cite{ref::xe100}, is used as a reference with its leakage fraction normalized to unity. An increase to $L_y=8.0$\,PE/keVee reduces the leakage fraction by almost a factor~4, the higher $L_y=12.0$\,PE/keVee by more than a factor~7. All other parameters which might affect background discrimination are kept constant here. 

\begin{figure}[tb]
\centering 
\includegraphics[width=.495\textwidth]{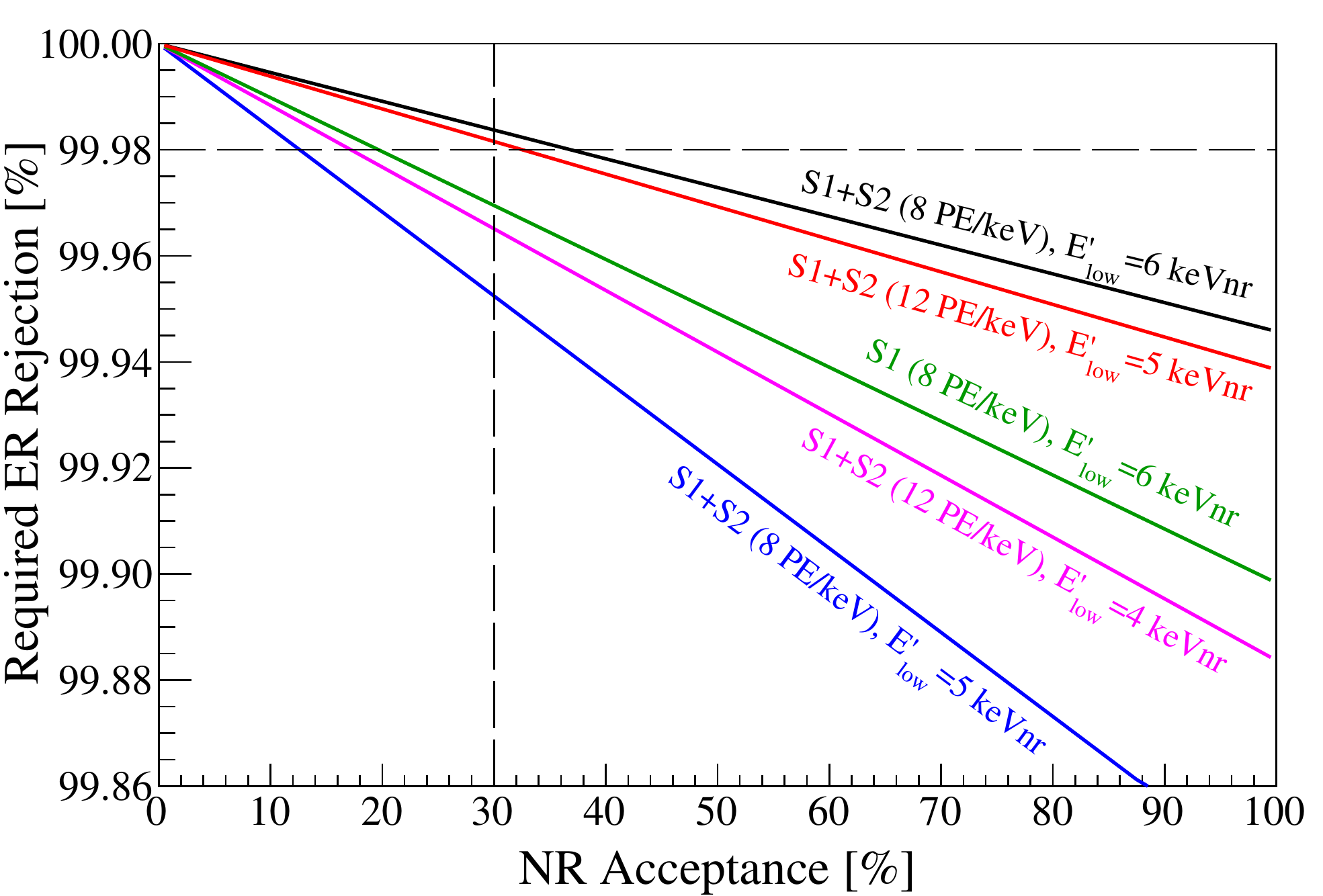}
\hfill
\includegraphics[width=.495\textwidth]{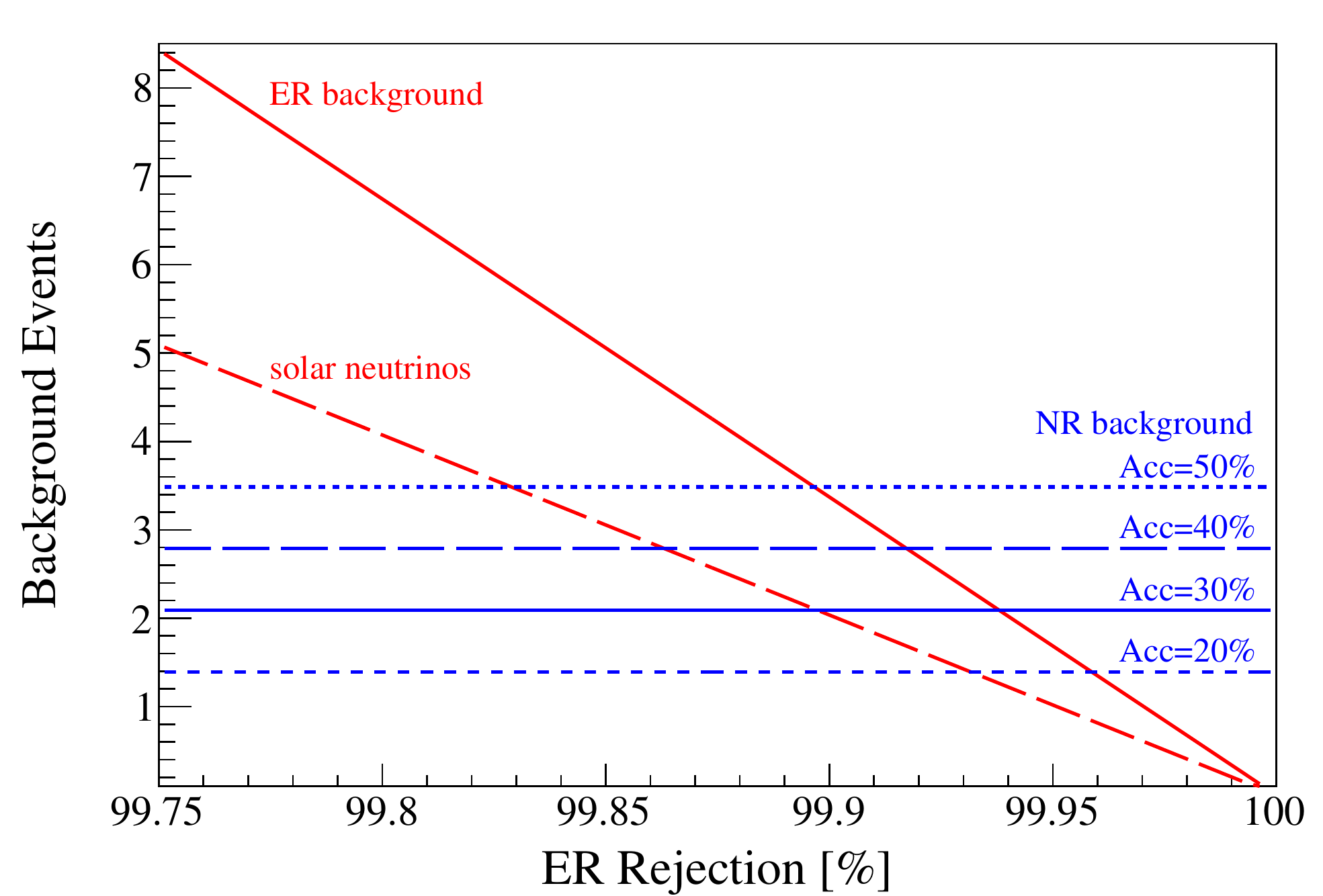}
\caption{{\bf (left)} ER rejection level required to achieve an ER background from solar pp- and $^7$Be neutrinos which is 20\% of the NR contribution from CNNS, as a function of NR acceptance. The comparison in a fixed low energy window ($X-20.5$\,keVnr) requires the choice of an energy scale and depends on the low energy threshold $E'_{low}$, due to the steeply rising CNNS spectrum. The figure shows five representative cases, the dashed lines indicate the ``reference'' numbers used in this study. {\bf (right)} Mean number of expected NR background events from CNNS and materials for different NR acceptances and mean number of expected ER background events (solar neutrinos, $^{85}$Kr, $^{222}$Rn, $2\nu\beta\beta$ and from materials as discussed in Section~\ref{sec::background}). The solar neutrinos are also shown independently. The NR background dominates for ER rejection levels better than 99.9\%. We assume an exposure of 200\,\ex, 5.0-20.5\,keVnr, and a combined energy scale with 8\,PE/keVee.
\label{fig::rejection} }
\end{figure}

In order to determine the ``typical'' ER rejection level required for a multi-ton scale LXe dark matter detector (and the corresponding NR acceptance), we compute the numbers at which the ER background from solar neutrinos is 20\% of the one from CNNS NRs in our low-energy window. An energy scale has to be selected in order to evaluate the ER background in the NR energy region. Due to the steeply rising CNNS spectrum, the rejection level depends crucially on the low energy threshold energy $E'_{low}$ (see also Section~\ref{sec::limit}); the upper limit is again fixed to $E'_{high} = 20.5$\,keVnr. For a variable NR acceptance, Figure~\ref{fig::rejection} (left) shows the required ER rejection level for five different combinations of energy scales, light yields, and threshold energies.
The resulting typical numbers are around 99.98\% ER rejection and 30\% NR acceptance (dashed lines), which we will further use as reference values. We note that the achievable WIMP sensitivity depends on the absolute size of the background, which varies by more than a factor~2 for the five cases in Figure~\ref{fig::rejection}, when evaluated at a fixed NR acceptance of 30\%. Figure~\ref{fig::rejection} (right) separately shows the mean expected number of events from all ER (neutrinos, 0.1\,ppt of $^\textnormal{\footnotesize nat}$Kr, 0.1\,$\mu$Bq/kg of $^{222}$Rn, materials) and NR sources (CNNS, materials), where the latter are given for different acceptances. The individual ER contribution of solar neutrinos is shown as well. We assume an exposure of 200\,\ex, a 5.0-20.5\,keVnr energy window and a combined energy scale with $L_y=8.0$\,PE/keVee. The NR background exceeds the one from ERs for rejection levels $\ge$99.9\%. As the background of an ultimate detector should be dominated by CNNS events, the ER rejection must be $\ge$99.98\% at a NR acceptance of 30-50\%.

\section{WIMP Sensitivity}
\label{sec::results}

\afterpage{\clearpage}

\begin{figure}[t]
\centering 
\includegraphics[width=.495\textwidth]{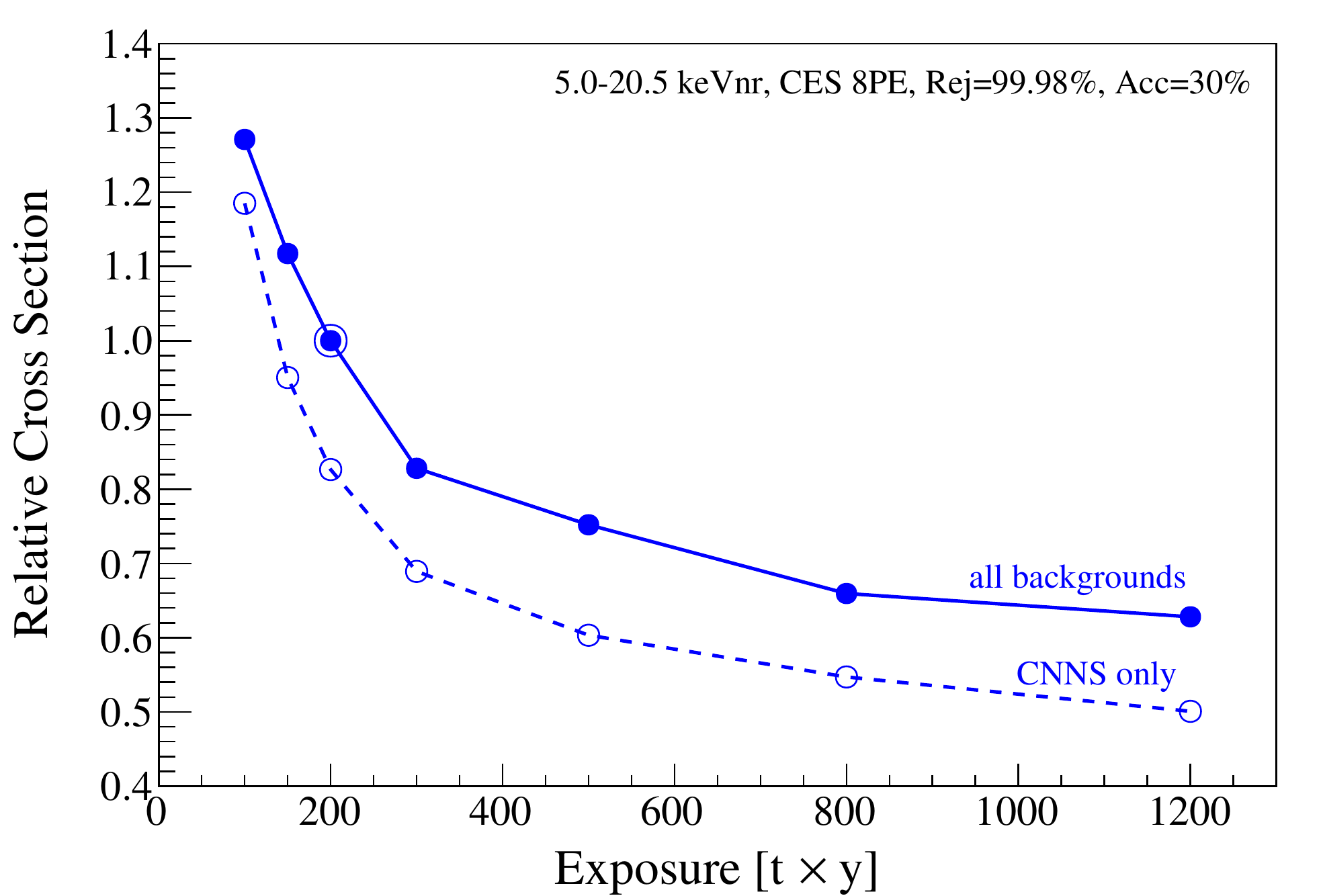}
\hfill
\includegraphics[width=.495\textwidth]{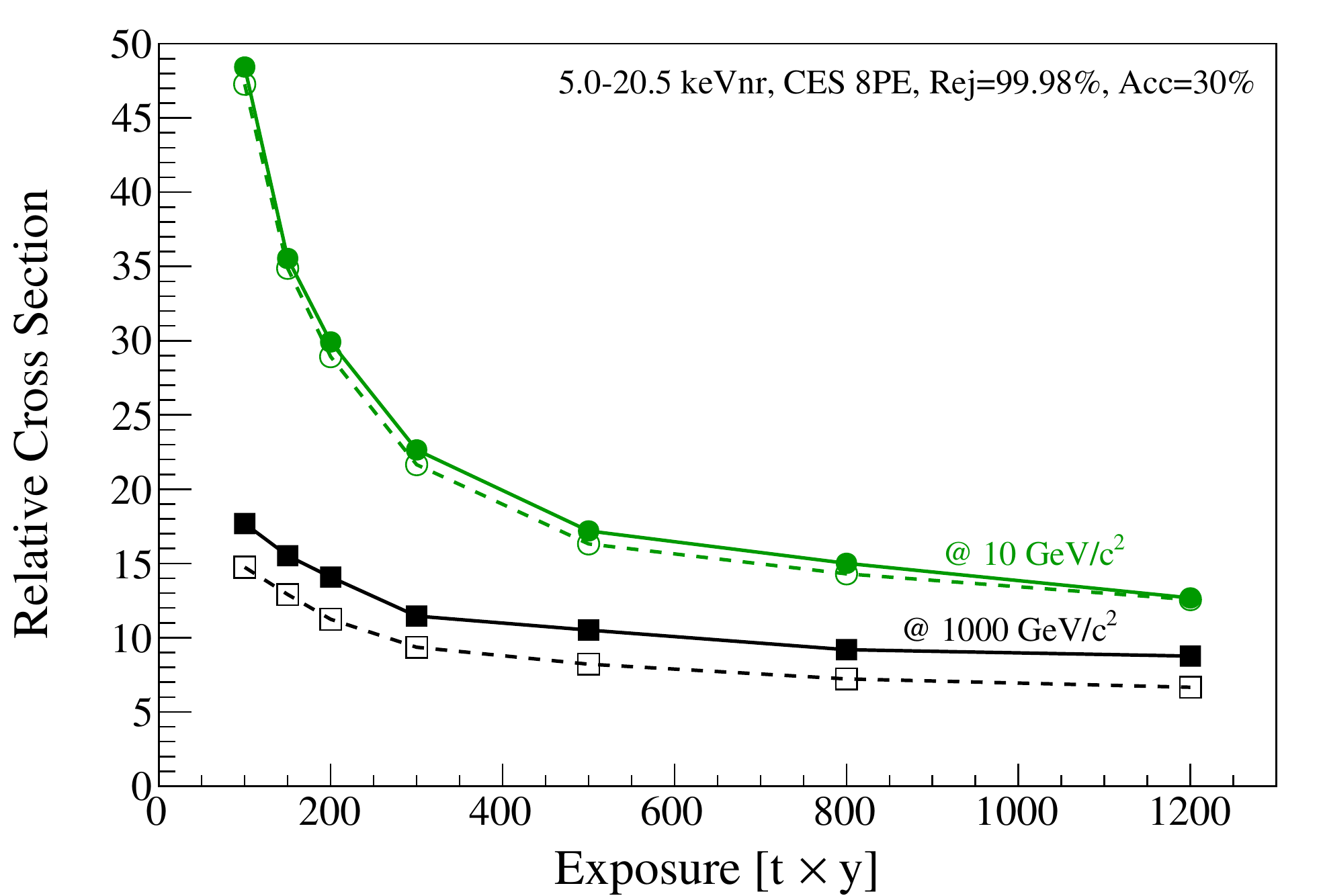}

\includegraphics[width=.495\textwidth]{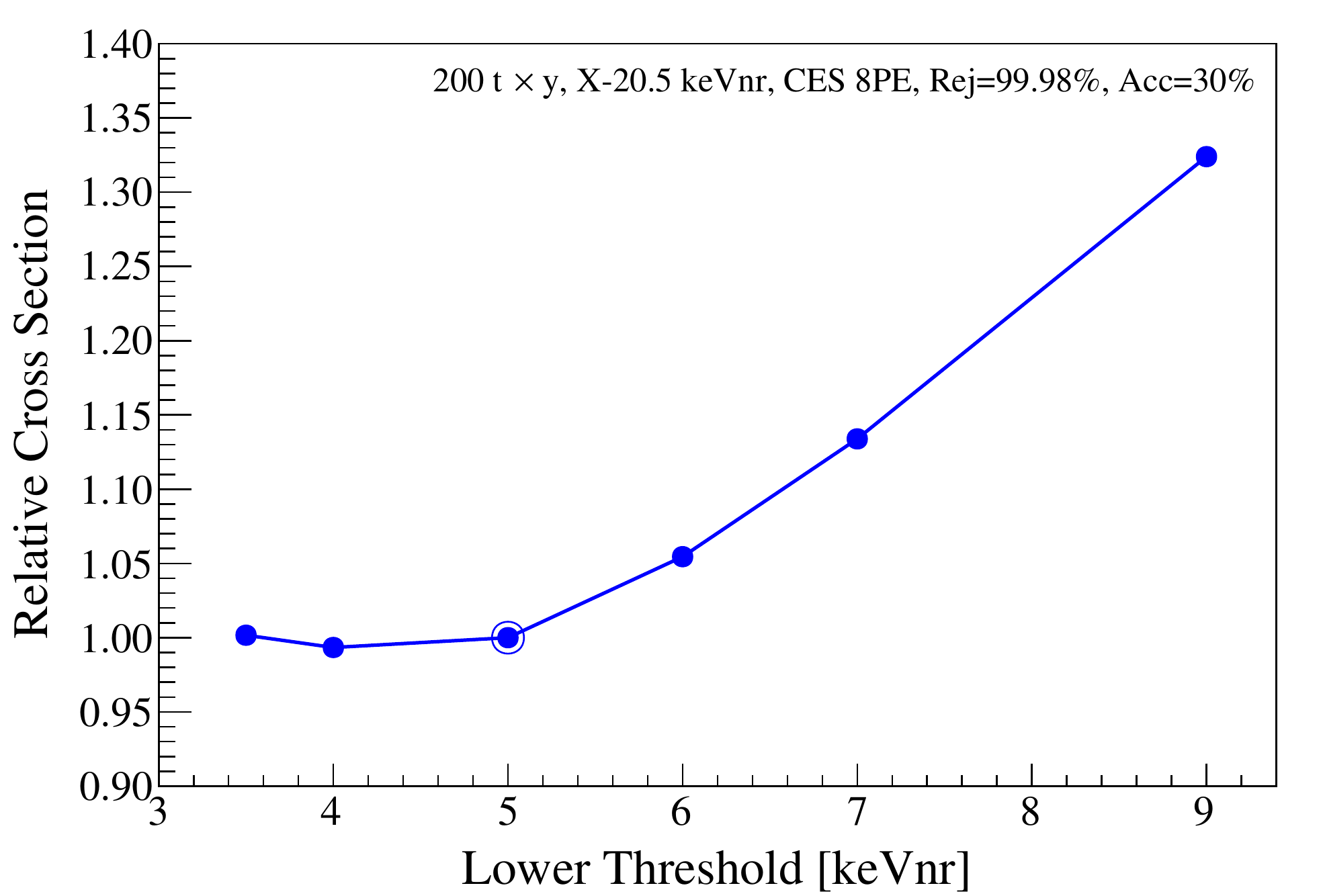}
\hfill
\includegraphics[width=.495\textwidth]{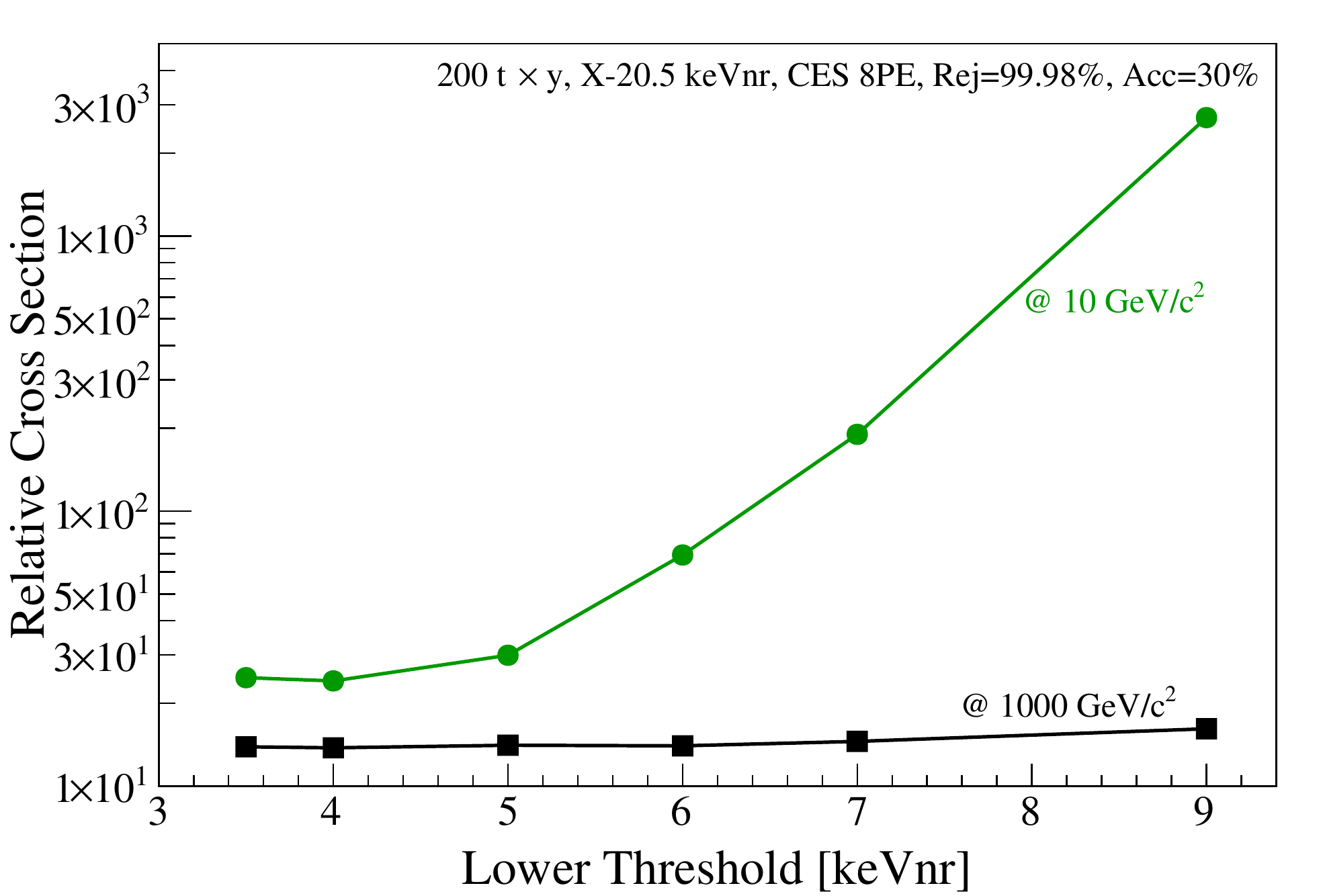}

\includegraphics[width=.495\textwidth]{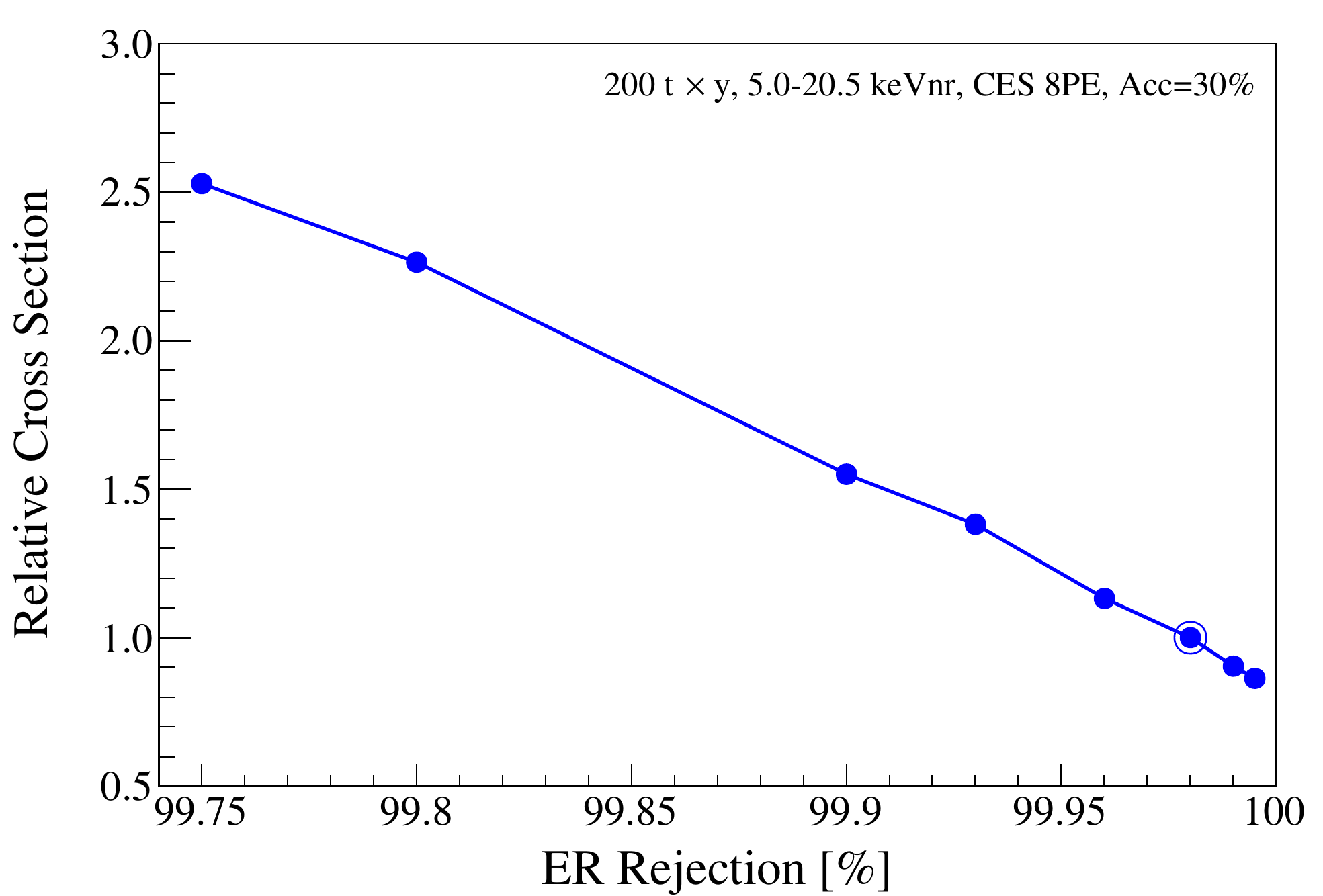}
\hfill
\includegraphics[width=.495\textwidth]{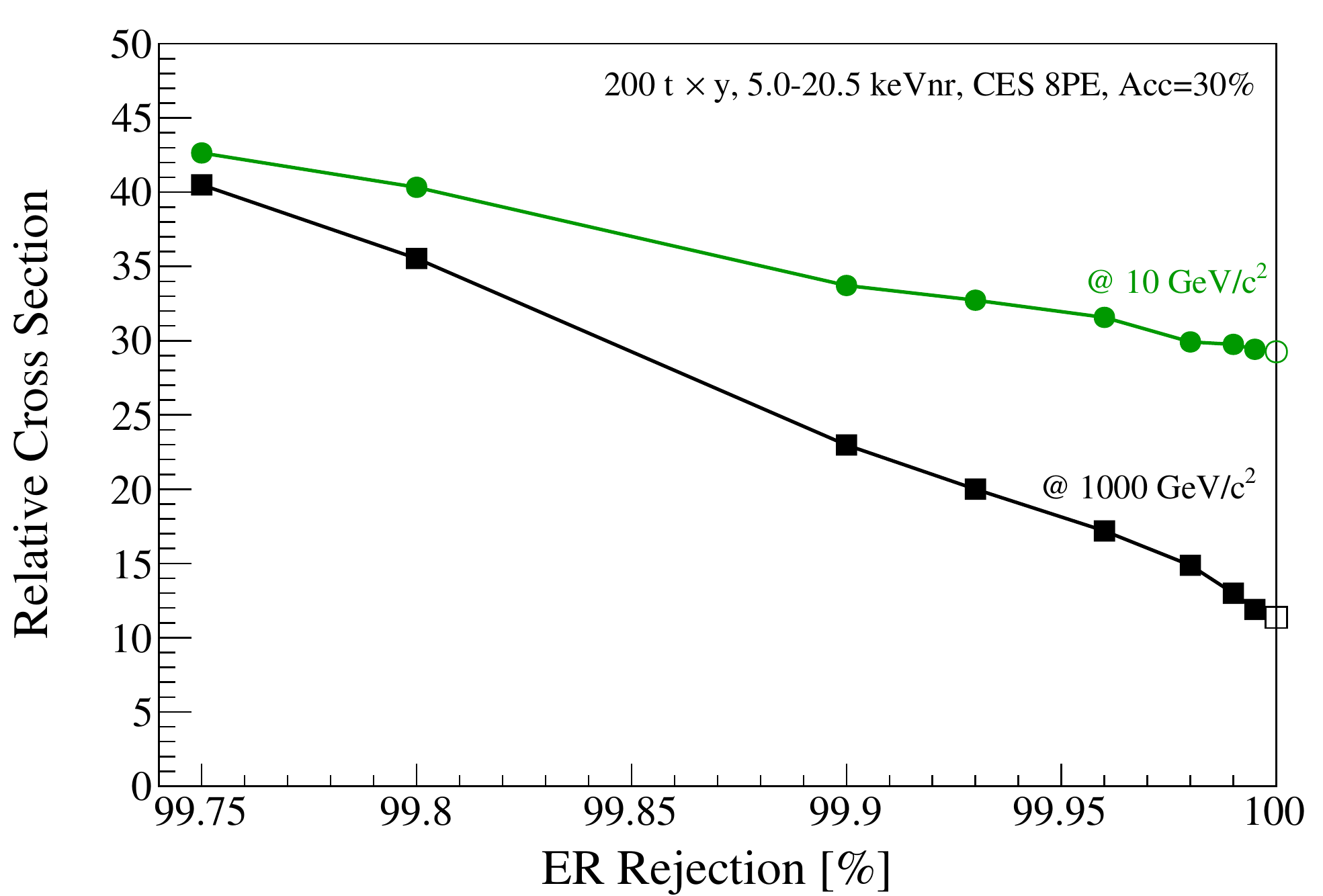}

\includegraphics[width=.495\textwidth]{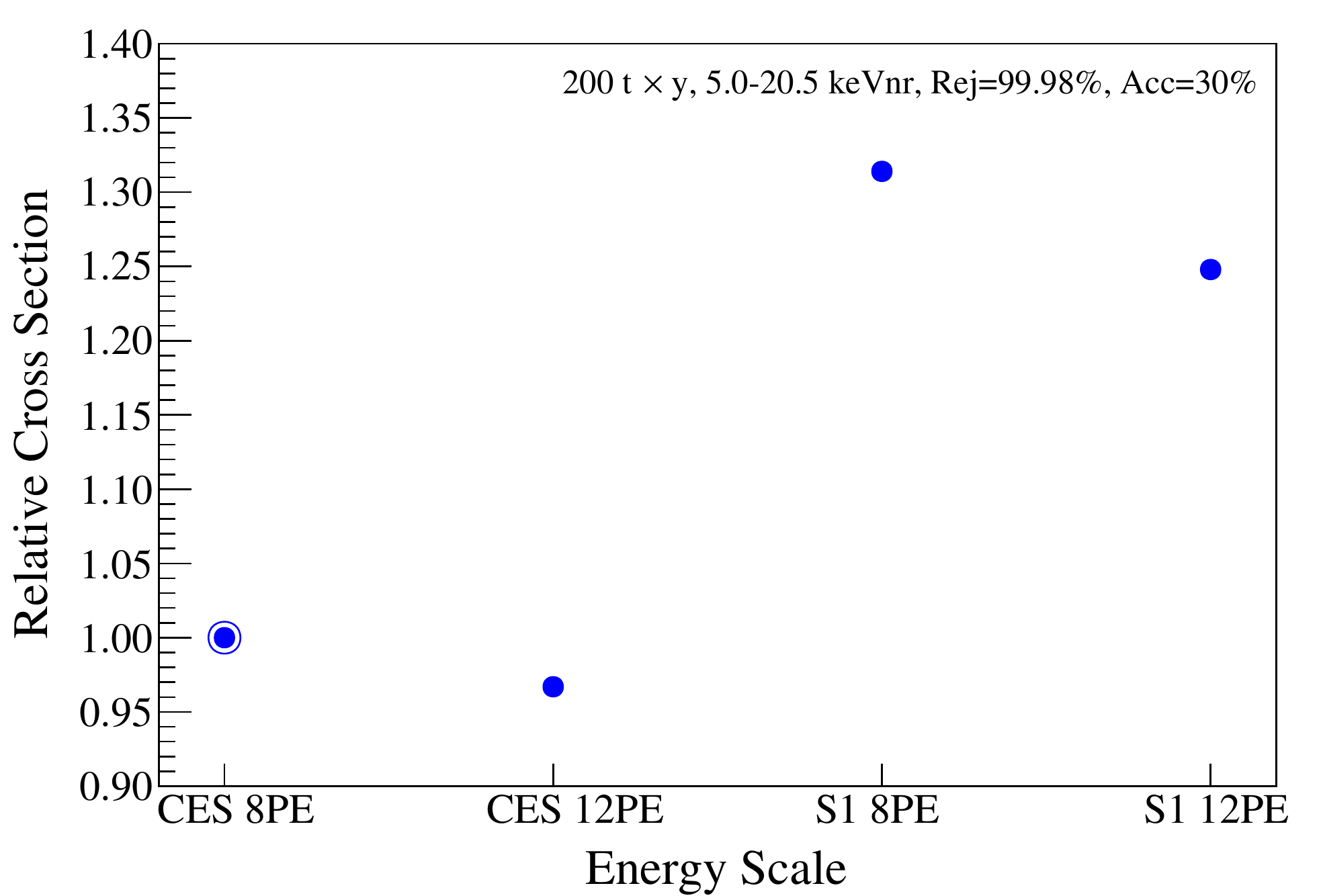}
\hfill
\includegraphics[width=.495\textwidth]{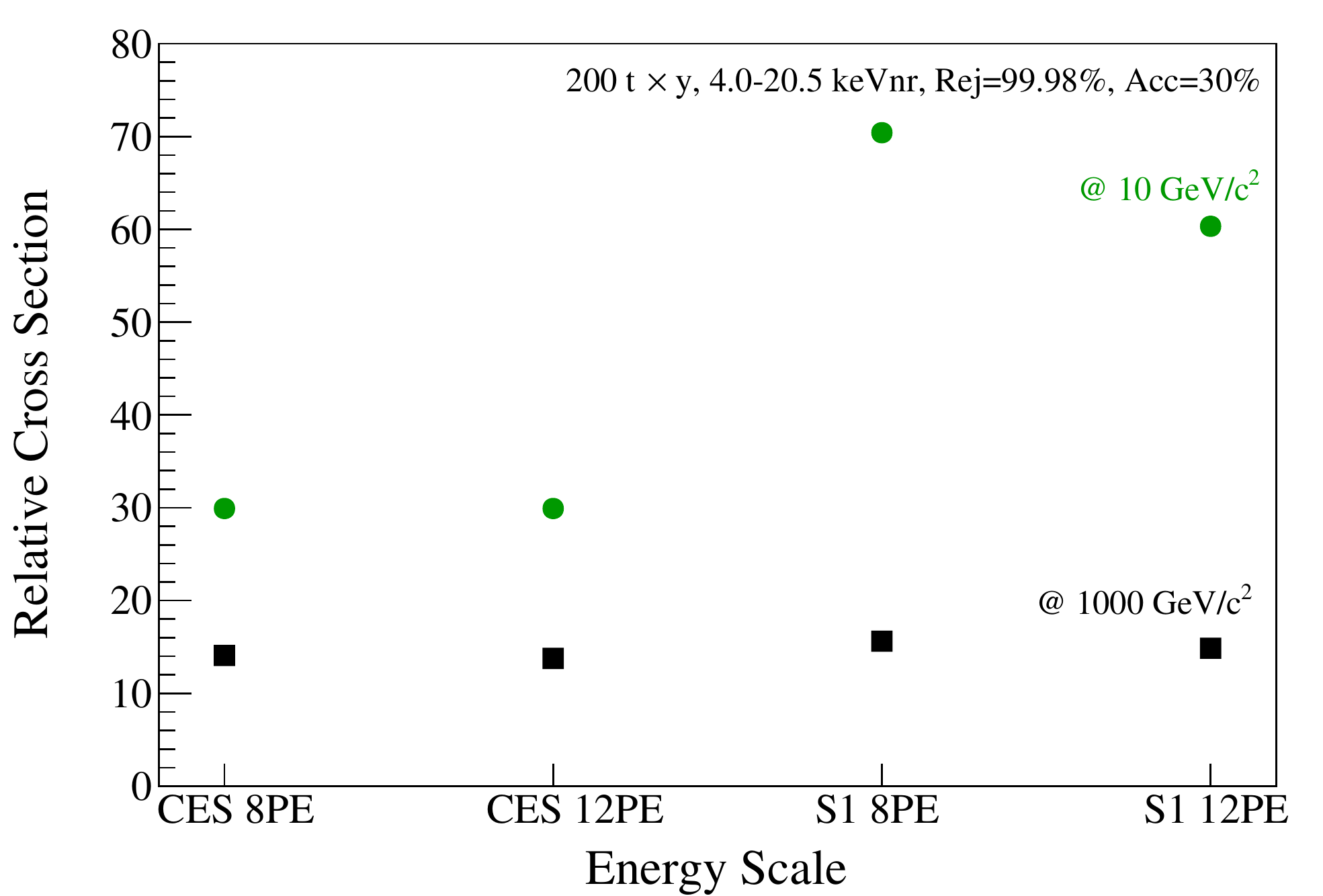}
\caption{Average sensitivity, defined as the 90\% CL exclusion limit on the cross section, at $\sim$40\,GeV/$c^2$ (left column) and at 10\,GeV/$c^2$ and 1000\,GeV/$c^2$ (right column). It is plotted vs.~(1\high{st} row) exposure, (2\high{nd}) lower threshold $E'_{low}$, (3\high{rd}) ER rejection level and (4\high{th}) energy scale. All values are normalized to the marked reference value, obtained by using ``standard'' parameters. The parameters not under study were fixed to the values indicated in the plots. Open symbols indicate the sensitivity considering only the CNNS background (at 30\% NR acceptance). See text for more discussion. \label{fig::eval} }
\end{figure}

The following experimental parameters were individually varied in a systematic way in order to evaluate their impact on the dark matter sensitivity:
\begin{itemize}
 \item exposure: target mass $M$ $\times$ live-time $T$,
 \item lower energy threshold $E'_{low}$,
 \item ER background rejection (at a fixed NR acceptance of 30\%), and
 \item energy scale: combined (S1+S2) and S1-only scale using two different light yields. 
\end{itemize}
The parameters not under study were fixed to the following values, which were found to be optimal for a realistic experiment: an exposure of 200\,\ex, a lower and an upper threshold of 5.0\,keVnr and 20.5\,keVnr, respectively, and an ER rejection level of 99.98\% ($2 \times 10^{-4}$). A combined energy scale (CES) with a light yield of $L_y=8$\,PE/keVee was used for the standard analysis and all background sources discussed in Section~\ref{sec::background} were taken into account.

The results are presented in Figure~\ref{fig::eval}. The four panels in the left column show the average sensitivity at the minimum of the sensitivity curve, typically around 40\,GeV/$c^2$, obtained for a fixed WIMP search box. The limit at 100\,GeV/$c^2$ is only about 60\% weaker, rendering this choice as typical for medium (weak scale) WIMP masses. The four panels on the right show the sensitivity at low ($m_\chi = 10$\,GeV/$c^2$) and high (1000\,GeV/$c^2$) WIMP masses. Fluctuations in the curves are of statistical nature. All sensitivities are given relative to the one obtained for the standard parameters, marked in the plots by the double ring.

\paragraph{Exposure} We study the WIMP sensitivity for exposures from 100\,\ex \ to 1200\,\ex \ (first row in Figure~\ref{fig::eval}). The CNNS background becomes relevant above 300\,\ex, where it contributes $\sim$3\,events to the background in the WIMP search box and the sensitivity starts to flatten. The gain in sensitivity is only a factor~1.3 when increasing the exposure from 300\,\ex \  to 1200\,\ex. 
At $m_\chi \sim 40$\,GeV/$c^2$, the improvement from the realistic background situation to the CNNS-only case (dashed lines, open markers) increases from a factor $\sim$1.05 at 100\,\ex \ to $\sim$1.25 at 1200\,\ex. With an increase from a factor~1.2 to~1.3, the situation is almost identical at 1000\,GeV/$c^2$. As the background is dominated by CNNS at low recoil energies, the difference is less pronounced at 10\,GeV/$c^2$, where the other background sources are basically irrelevant, such that the two sensitivities almost agree at 10\,GeV/$c^2$. 
The gain in sensitivity for the CNNS-only case shown here, assuming no ER background but a finite NR efficiency of 30\%, is less pronounced than in Figure~\ref{fig::cnns} (right), where 100\% NR was assumed to illustrate the most optimistic background scenario.

\paragraph{Lower energy threshold} Changing the lower energy threshold $E'_{low}$ has the largest impact at small WIMP masses, due to their expected steeply falling recoil spectrum. This is shown in the second row of Figure~\ref{fig::eval}, where the sensitivity at 10\,GeV/$c^2$ decreases by two orders of magnitude when increasing the lower threshold from 3.5\,keVnr to 9\,keVnr in a 200\,\ex \ run. Around $E'_{low}\approx5$\,keVnr, the sensitivity starts to level off, and there is no significant sensitivity gain by going to lower thresholds, as the irreducible CNNS background effectively builds a ``wall'' of background events, limiting the energy range with an acceptable signal-to-background ratio. The sensitivity changes more moderately (15-30\%) for medium and high WIMP masses due to the flatter expected recoil spectra.

\paragraph{Discrimination level} The presence of ER background requires a sizeable ER rejection factor, see also the discussion in Section~\ref{sec::rejection}. In the third row of Figure~\ref{fig::eval}, we show how the sensitivity (exposure 200\,\ex) depends on the discrimination level, which is varied from 99.75\% ($2.5\times10^{-3}$ ER acceptance) to 99.995\% ($5\times10^{-5}$). For all WIMP masses, it increases linearly with the increasing rejection level. At 99.98\% rejection, it is only 20\% higher compared to the CNNS-only case (30\% NR acceptance, open symbols) for medium and high WIMP masses. At low masses, however, due to the dominating CNNS background, the slope of the improvement is  weaker and the sensitivity flattens already at rejection levels $>$99.9\%. The fluctuations in the curves are statistical in nature.

\paragraph{Energy scale} As already discussed and illustrated in Figure~\ref{fig::cnns} (left), the choice of energy scale is important in order to achieve the best possible sensitivity. This is quantified in the last row of Figure~\ref{fig::eval}, comparing four different energy scales: two employ a combined energy scale (CES), using the light (S1) and the charge information (S2) simultaneously, and two rely on the S1 signal only, see also Section~\ref{sec::escale}. Two different light yields at 122\,keVee, $L_y=8$\,PE/keVee and $L_y=12$\,PE/keVee, are used. The better resolution of the combined energy scale leads to significantly increased sensitivities at low WIMP masses, but also improves the sensitivity for medium and -- to a much lesser extent -- high masses. It is interesting to note that the 50\% increase in $L_y$ has an almost negligible impact at fixed ER rejection/NR acceptance. Only at 10\,GeV/$c^2$ when using the S1-only scale, the improved resolution due to the higher $L_y$ helps to deal with the CNNS background. In general, the increased light yield is expected to improve the ER rejection and therefore the sensitivity, as discussed in Section~\ref{sec::rejection} and above.

At a given $E'_{low}$ the S1-only scale is somewhat more sensitive to very low WIMP masses ($m_\chi \le 7$\,GeV/$c^2$) than the CES scale. This is because of its lower energy resolution, which allows probing the tail of the WIMP recoil spectrum by recording upward fluctuations, at cross sections where the CNNS background is not yet dominating. We note, however, that the quantification of this effect depends crucially on the relative scintillation efficiency \leff \ employed for the study.
\smallskip

A likelihood analysis was employed in order evaluate a realistic WIMP sensitivity for an exposure of 200\,\ex \ and a 5-35\,keVnr energy interval. In energy space, ER and NR signals were distributed using a combined energy scale with $L_y=8.0$\,PE/keVee, and in discrimination space realizing a 99.98\%~ER rejection at 30\%~NR acceptance. However, the rejection level was not fixed in the analysis. If no signal is observed, such an experiment will have the sensitivity to exclude spin-independent WIMP-nucleon scattering cross sections of $2.5 \times 10^{-49}$\,cm$^2$ for 40\,GeV/$c^2$ WIMPs ($3.1 \times 10^{-48}$\,cm$^2$ at 1000\,GeV/$c^2$), as shown in Figure~\ref{fig::size} (left). 
It improves upon the expected sensitivities of the XENONnT and LZ projects (exposure $\sim$20\,\ex) by an order of magnitude, covering most of the experimentally accessible parameter space. It approaches the ``discovery limit'' suggested in Ref.~\cite{ref::billard}, which is reached with an increased exposure of 500\,\ex.

\section{Summary and Discussion}
\label{sec::discussion}

\begin{figure}[t!]
\centering 
\includegraphics[width=.495\textwidth]{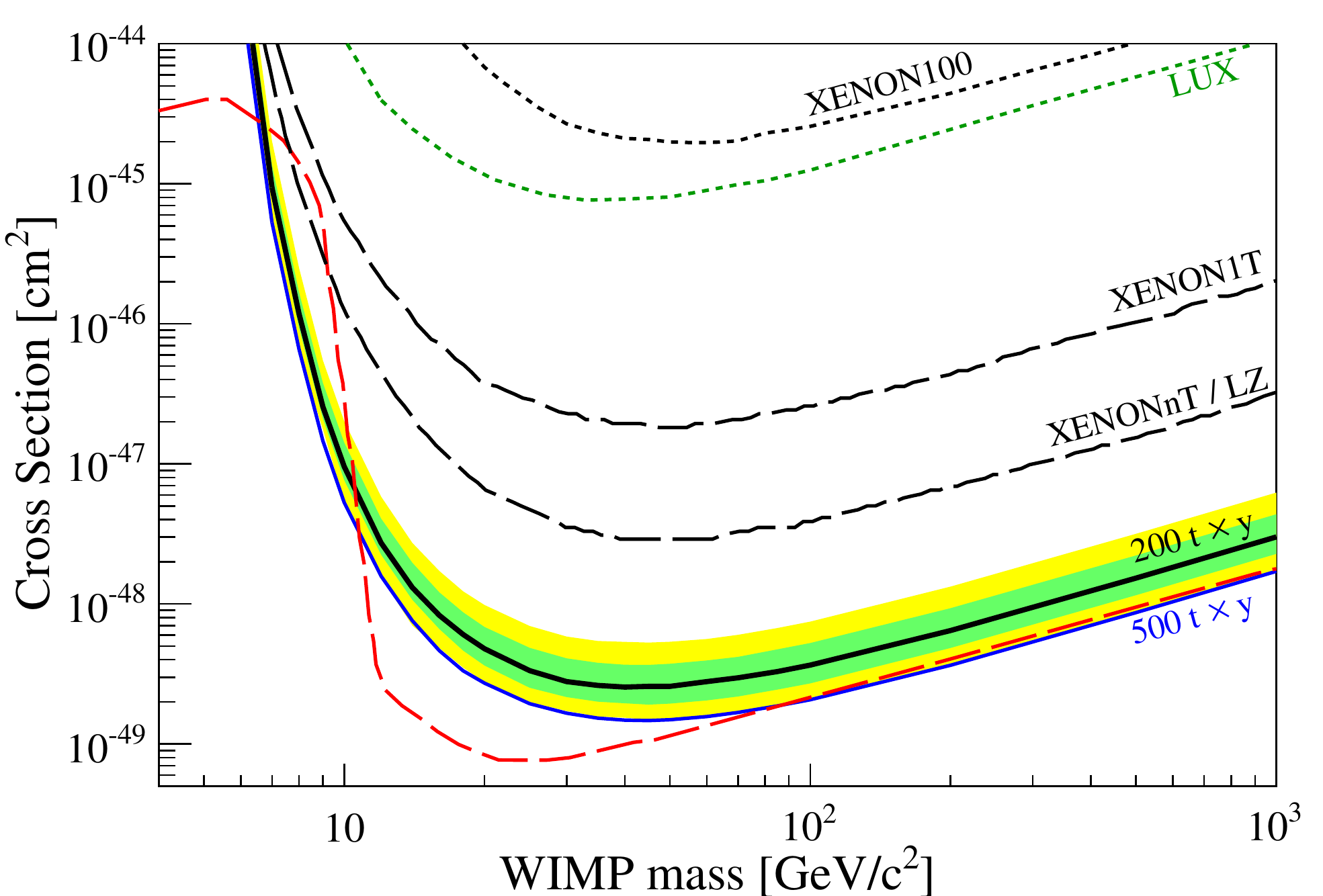}
\hfill
\includegraphics[width=.495\textwidth]{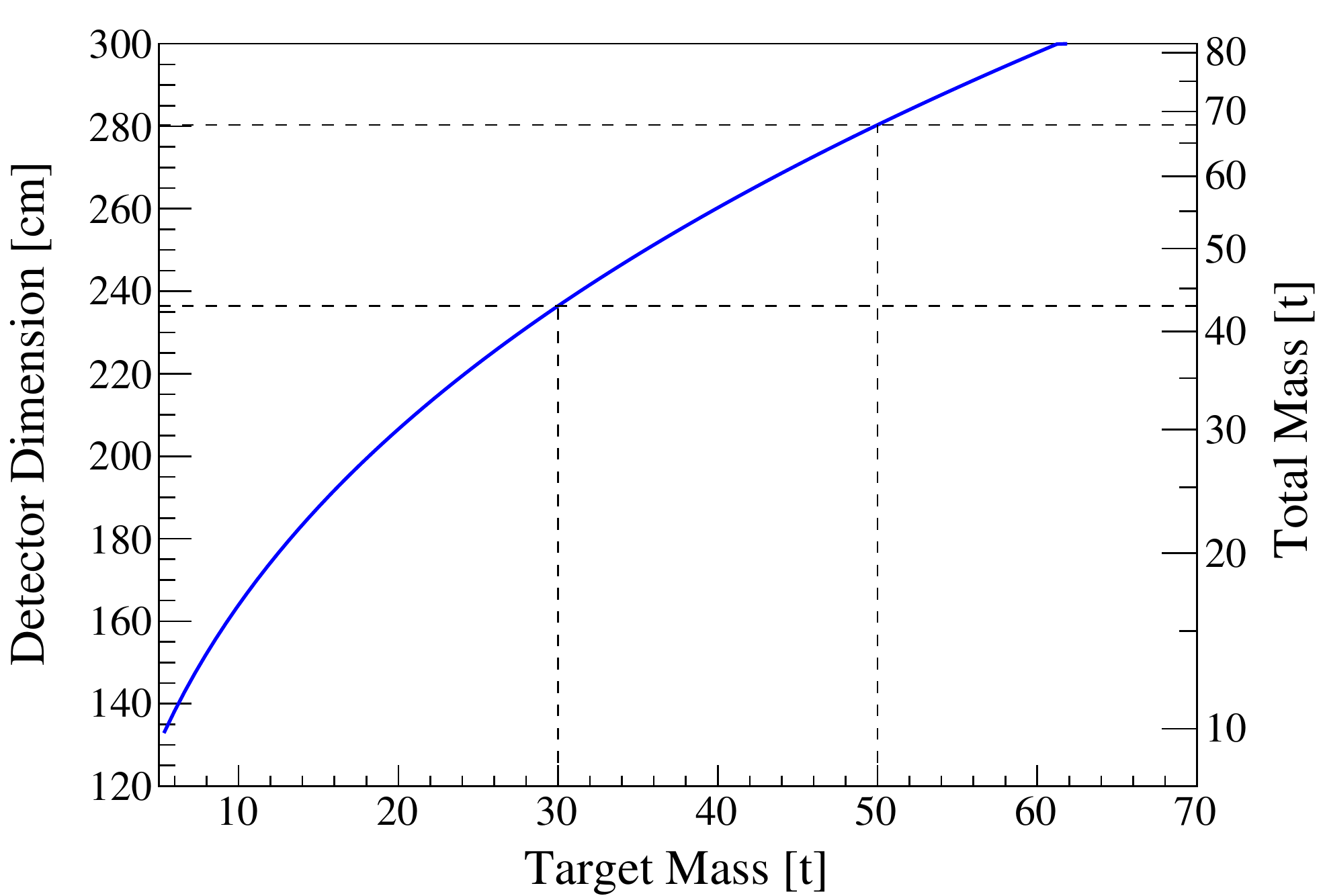}
\caption{{\bf (left)} Estimated sensitivity reach of a multi-ton LXe detector to spin-independent WIMP-nucleon scattering cross sections for exposures of 200\,\ex \ (black, with $1\sigma$/$2\sigma$ intervals) and 500\,\ex \ (blue). It was derived using a likelihood analysis assuming a 5-35\,keVnr energy interval, a separation of the ER and NR distributions corresponding to 99.98\% ER rejection at 30\% NR acceptance, and a combined energy scale with $L_y=8.0$\,PE/keVee. The results are compared to published limits from XENON100~\cite{ref::xe100run10} and LUX~\cite{ref::luxresult}, and to the reach of the upcoming projects XENON1T~\cite{ref::xe1t}, XENONnT~\cite{ref::xe1t_mv} and LZ~\cite{ref::lz}. The ``WIMP discovery limit'' of Ref.~\cite{ref::billard} is also shown (red dashed).  {\bf (right)} Estimate of the typical linear dimension (diameter and height) of a cylindrical LXe time projection chamber vs.~the target mass used for the WIMP search. The reduction of external backgrounds requires a larger detector scale (we assume extra 30\,cm) and more LXe (right scale). \label{fig::size} }
\end{figure}

We have studied the sensitivity of a DARWIN-like multi-ton scale LXe detector to spin-independent WIMP-nucleon interactions by means of toy Monte Carlo experiments, which take into account all known backgrounds and a realistic detector response. The simulated outcomes have been analyzed using a WIMP search box and Yellin's maximum gap method~\cite{ref::yellin}, and by means of an unconstrained likelihood analysis for selected cases. We have examined the sensitivity dependence on several experimental parameters and can draw the following conclusions in order to optimize the dark matter reach:
\begin{itemize}
 \item A minimal exposure of 100\,\ex \ is required to probe cross sections of a few $\times$10$^{-49}$\,cm$^2$ at $\sim$40\,GeV/$c^2$, assuming realistic detector parameters. Sensitivities as low as $2.5 \times 10^{-49}$\,cm$^2$ can be reached with 200\,\ex \ (see Figure~\ref{fig::size}, left).
 \item An energy threshold of 5\,keVnr or below must be achieved in order keep sensitivity to WIMP masses below $\sim$10\,GeV/$c^2$. However, the CNNS-induced background increases significantly at lower energies.
 \item The achievable sensitivity crucially depends on the rejection efficiency for ER backgrounds, which mainly stem from low energy solar neutrinos. The goal is 99.98\% rejection or better. Such a discrimination level has already been demonstrated by LXe experiments~\cite{ref::zeplin3,ref::kaixuan}. The mean expected sensitivity at this level, taking into account all backgrounds, is only $\sim$20\% worse than the (unrealistic) case in which all sources besides the one from CNNS can be rejected completely. This factor is only mildly dependent on the assumed exposure and increases from $10\%$ to 30\% from 100\,\ex \ to 1200\,\ex. 
 \item The concentration of $^\textnormal{\footnotesize nat}$Kr has to be reduced to a level around 0.1\,ppt to achieve a background level which is a factor $\sim$2.5 below the one from solar neutrinos~\cite{ref::darwinnu}. The concentration of $^{222}$Rn has to be $<$0.3\,$\mu$Bq per kg of LXe target in order to contribute to the background at the same level as $^{85}$Kr. With these assumptions, the background is dominated by NR events from CNNS interactions at rejection levels around 99.98\%.
 \item The NR energy scale should be reconstructed using the light and charge signals. The improved energy resolution of such a scale is crucial in dealing with the CNNS events induced by $^8$B neutrinos.
 \item A high light yield is necessary to establish the required ER rejection levels, and we use $L_y=8.0$\,PE/keVee for most of the study. However, a higher $L_y$ does not significantly improve the sensitivity reach at a given rejection/acceptance level. Light yields of this magnitude have been achieved already, also in large detectors, e.g., LUX~\cite{ref::lux}.
\end{itemize}

The experiment must be operated over a time span of a few years, therefore defining the mass-scale of the future instrument. A fiducial target of 30\,t of LXe would require a total of 6.7\,y of science data to collect 200\,\ex, plus additional time for detector calibration. As illustrated in Figure~\ref{fig::size} (right), such a LXe target has a typical dimension of 235\,cm, assuming a cylindrical geometry with equal diameter and height. In order to realize the required self-shielding to achieve the external background level discussed in Section~\ref{sec::background}, an additional LXe layer of $\sim$15\,cm has to be added all around the target. This requires a total LXe mass of 43\,t and an inner cryostat vessel with a typical dimension of $\sim$265\,cm. LXe TPCs of this size have not yet been realized, and further R\&D will be necessary in order to overcome experimental challenges such as ultra-high voltage, target purity, high light and charge yields, charge drift, calibration and low backgrounds, especially in terms of $^{85}$Kr and $^{222}$Rn.

With its excellent sensitivity, a DARWIN-type multi-ton LXe detector will be a valuable tool to further study the WIMP particle, should it be detected in the upcoming generation of experiments~\cite{ref::bigdmdet}. A  $m_\chi=500$\,GeV/$c^2$ (1000\,GeV/$c^2$) WIMP with a cross section around $2 \times 10^{-47}$\,cm$^2$, as predicted by Ref.~\cite{ref::wimpqcd} for wino-type dark matter and at the sensitivity limit of XENON1T~\cite{ref::xe1t}, would produce 46~(24)~signal events in a 5-35\,keVnr, 30\% NR acceptance WIMP search window of a DARWIN detector with the sensitivity shown in Figure~\ref{fig::size} (left, 200\,\ex). Below $m_\chi < 200$\,GeV/$c^2$, LXe detectors are particularly well-suited to reconstruct the main WIMP parameters, mass and cross section, from the observed recoil spectrum~\cite{ref::newstead}. For $m_\chi=30$\,GeV/$c^2$ and $m_\chi=100$\,GeV/$c^2$, with cross sections of $2 \times 10^{-48}$\,cm$^2$, DARWIN would detect 22~and 19~events from spin-independent interactions in the WIMP search region, respectively. We note that for WIMP masses above $\sim$200-250\,GeV/$c^2$, the WIMP mass cannot be fully reconstructed by any target material, due to the degeneracy in $\rho_0 \ \sigma \ m_\chi^{-1}$ for massive WIMPs with $m_\chi \gg m_{Xe}$~\cite{ref::bigdmdet}.

\begin{figure}[tbp]
\begin{minipage}[]{0.49\textwidth}
\begin{center}
\includegraphics[width=1.0\textwidth]{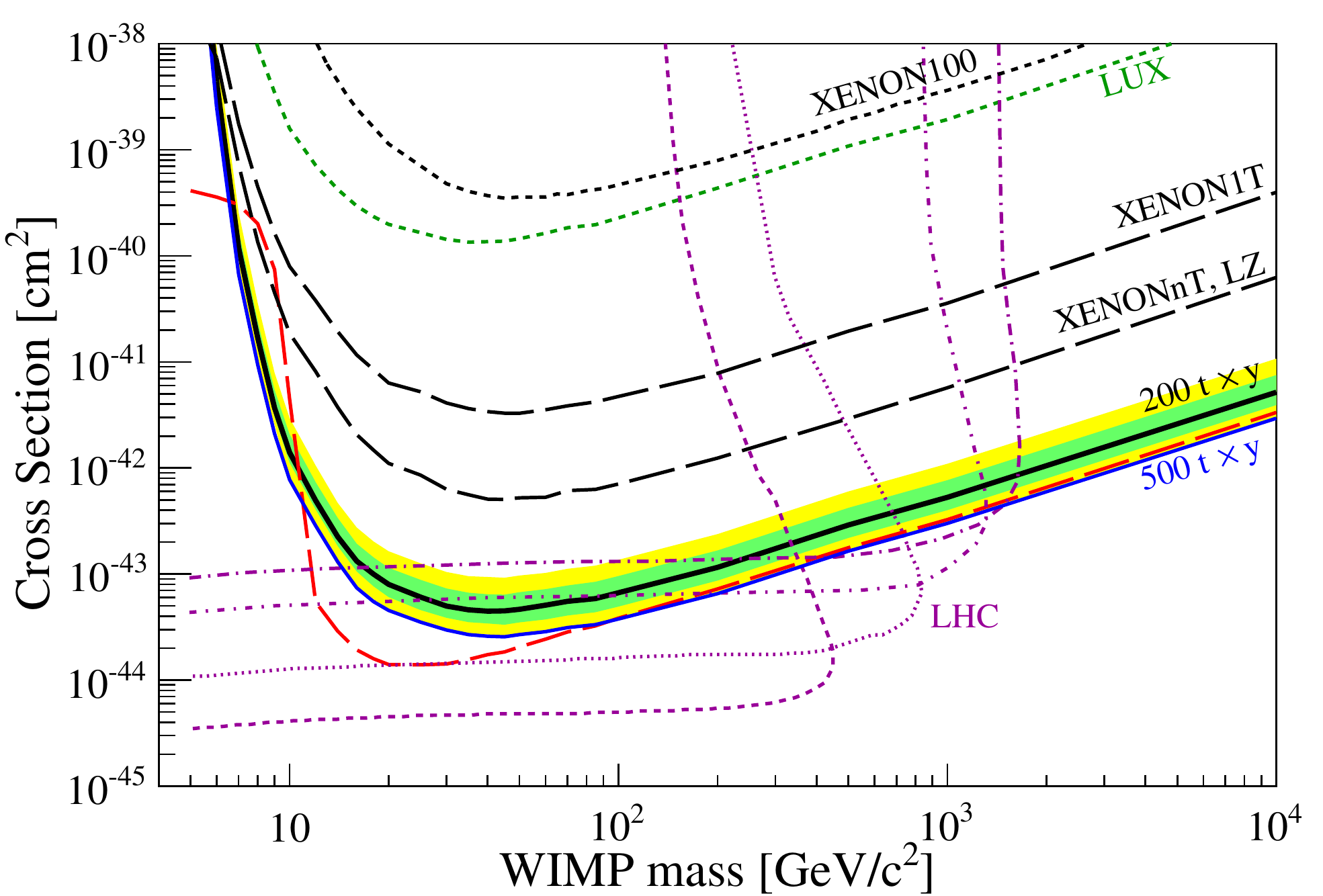}
\end{center}
\end{minipage}
\hfill
\begin{minipage}[]{0.49\textwidth}
\caption{Sensitivity of a DARWIN-type LXe detector to spin-{\it de}pendent WIMP-nucleon cross sections, assuming neutron-only couplings. Results from a likelihood analysis for 200\,\ex \ and 500\,\ex \ exposures are shown (assumptions as in Figure~\ref{fig::size}, left), together with the limit by XENON100~\cite{ref::xe100sd}, an interpretation of the LUX data~\cite{ref::savage_lux} and estimated sensitivities for XENON1T, XENONnT, and LZ. DARWIN and the high-luminosity LHC will cover common parameter space. The 14\,TeV LHC limits for the coupling constants $g_\chi$\,$=$\,$g_q$\,$=$\,0.25, 0.5, 1.0, 1.45 (bottom to top) are taken from~\cite{ref::lhc_comp}.    \label{fig::spindep} }
\end{minipage}
\end{figure}

Probing identical or largely overlapping regions of parameter space is mandatory in order to exploit the complementarity of the different approaches to detect dark matter, with the goal of identifying the nature of the WIMP and studying its quantum numbers. Especially for spin-{\it de}pendent interactions, a DAR\-WIN-type LXe detector will largely cover parameter space which is also probed by the high-luminosity LHC, operating at 14\,TeV center-of-mass energy. This is shown in Figure~\ref{fig::spindep}, where we compare the LHC mono-jet sensitivity to spin-dependent interactions (neutron-only couplings)~\cite{ref::lhc_comp} with two multi-ton LXe exposures of 200\,\ex \ and 500\,\ex \ (5-35\,keVnr, CES with $L_y=8.0$\,PE/keVee, likelihood sensitivity for 99.98\% ER rejection at 30\% NR acceptance). The LHC 90\% CL exclusion limits, reaching up to $m_\chi \approx 1$\,TeV/$c^2$, were calculated for four cases, each time assuming the same values for the coupling constants to quarks $g_q$ and the Dirac fermion WIMP $g_\chi$. 
The expected sensitivities for a DARWIN detector as well as for XENON1T, XENONnT and LZ were directly derived from the spin-independent sensitivities $\sigma_{SI}(m_\chi)$ shown in Figure~\ref{fig::size} (left). A scaling factor $f(m_\chi)$ was obtained by comparing the spin-independent~\cite{ref::xe100run10} and the spin-dependent results~\cite{ref::xe100sd} from XENON100, which are based on exactly the same dataset. The spin-dependent limits are then given by $\sigma_{SD}(m_\chi) = f(m_\chi) \times \sigma_{SI}(m_\chi)$. For LUX, we obtain the identical exclusion curve as published by~\cite{ref::savage_lux} for $m_\chi \ge 10$\,GeV/$c^2$ (and somewhat weaker limits at lower masses), confirming the validity of our approach. 

In general, the LHC will only be able to probe WIMPs up to $\sim$1\,TeV even at 14\,TeV. For spin-{\it in}dependent couplings, shown in Figure~\ref{fig::size} (left), the science reach of direct detection detectors is many orders of magnitude superior to future LHC searches, which will not reach below $10^{-44}$\,cm$^2$~\cite{ref::lhc_comp}. Therefore, direct searches -- such as multi-ton LXe detectors covering the entire mass-space above $\sim$6\,GeV/$c^2$ -- are necessary even at low WIMP masses to detect dark matter if it couples to matter via spin-independent interactions.

\acknowledgments 
This work was supported by the Swiss National Science Foundation (SNSF) via grant numbers~149445 and~149256, the Albert Einstein Center for Fundamental Physics (AEC) at the University of Bern, the University of Z\"urich and the INFN.

\end{document}